\documentclass[]{aastex631}
\usepackage{amsmath}

\received{}
\revised{}
\accepted{}

\submitjournal{The Astrophysical Journal}

\shorttitle{Relativistic Hydrodynamic Code}
\shortauthors{Seo et al.}

\begin{document}

\title{A Simulation Study of Ultra-Relativistic Jets - I. A New Code for Relativistic Hydrodynamics}
\author[0000-0002-5550-8667]{Jeongbhin Seo}
\affiliation{Department of Earth Sciences, Pusan National University, Busan 46241, Korea}
\author[0000-0002-4674-5687]{Hyesung Kang}
\affiliation{Department of Earth Sciences, Pusan National University, Busan 46241, Korea}
\author[0000-0002-5455-2957]{Dongsu Ryu}
\affiliation{Department of Physics, College of Natural Sciences, UNIST, Ulsan 44919, Korea}
\author[0000-0002-0786-7307]{Seungwoo Ha}
\affiliation{Department of Physics, College of Natural Sciences, UNIST, Ulsan 44919, Korea}
\author[0000-0002-2133-9324]{Indranil Chattopadhyay}
\affiliation{Aryabhatta Research Institute of Observational Sciences, Nainital, India}
\correspondingauthor{Dongsu Ryu}
\email{ryu@sirius.unist.ac.kr}

\begin{abstract}


In an attempt to investigate the structures of ultra-relativistic jets injected into the intracluster medium (ICM) and the associated flow dynamics, such as shocks, velocity shear, and turbulence, we have developed a new special relativistic hydrodynamic (RHD) code in the Cartesian coordinates, based on the weighted essentially non-oscillatory (WENO) scheme. It is a finite difference scheme of high spatial accuracy, which has been widely employed for solving hyperbolic systems of conservation equations. The code is equipped with different WENO versions, such as the 5th-order accurate WENO-JS \citep{jiang1996}, WENO-Z, and WENO-ZA, and different time integration methods, such as the 4th-order accurate Runge-Kutta (RK4) and strong stability preserving RK (SSPRK), as well as the implementation of the equations of state (EOSs) that closely approximate the EOS of the single-component perfect gas in relativistic regime. In addition, it incorporates a high-order accurate averaging of fluxes along the transverse directions to enhance the accuracy of multi-dimensional problems, and a modification of eigenvalues for the acoustic modes to effectively control the carbuncle instability. Through extensive numerical tests, we assess the accuracy and robustness of the code, and choose WENO-Z, SSPRK, and the EOS suggested in \citet{ryu2006} as the fiducial setup for simulations of ultra-relativistic jets. The results of our study of ultra-relativistic jets using the code is reported in an accompanying paper \citep[][Paper II]{seo2021b}.

\end{abstract}


\section{Introduction}
\label{s1}

High-energy astrophysical phenomena, such as pulsar wind nebulae, gamma-ray bursts (GRBs), and radio-loud active galactic nuclei (AGNs), often involve relativistic jets and flows \citep[see, e.g.,][for reviews]{gaensler2006,piran2005, hardcastle2020}. Relativistically beamed jets from blazers are inferred to have the bulk Lorentz factor of up to $\Gamma \sim 50$ \citep[e.g.,][]{Savolainen2010,lister2013}. In the most widely accepted model for GRBs, a highly focused explosion associated with the formation of a blackhole drives a pair of collimated relativistic jets with $\Gamma \lesssim 400$ \citep[e.g.,][]{nava2017}. 

For studies of relativistic hydrodynamic (RHD) problems, numerical codes have been built by adopting various schemes originally developed for non-relativistic Newtonian hydrodynamics. A partial list of RHD codes includes the followings: \citet{falle1996} based on the van Leer scheme, \citet{marti1996}, \citet{aloy1999}, and \citet{mignone2005b} based on the PPM scheme, \citet{zhang2006}, \citet{radice2012}, and \citet{nunez2016} implementing high-order reconstructions including WENO, \citet{schneider1993} and \citet{mignone2005a} based on the HLLE and HLLC schemes, \citet{ryu2006} based on the TVD scheme, \citet{dolezal1995} based on the ENO scheme, \citet{del2002} based on the CENO scheme, \citet{qin2016} based on the DG scheme, and \citet{duffell2011} and \citet{he2012} based on moving-mesh methods. Reviews of RHD codes can be found, for instance, in \citet{marti2003, marti2015}.

In an accompanying paper \citep[][Paper II]{seo2021b}, we report the findings of our recent study on the structures of ultra-relativistic jets injected into the intracluster medium (ICM) and the associated flow dynamics, such as shocks, velocity shear, and turbulence, through high-resolution RHD simulations. In this paper, we describe the special RHD code newly developed for that study. Our code is based on the weighted essentially non-oscillatory (WENO) scheme for hyperbolic systems of conservation equations, and high-order accurate time integration methods. It also incorporates a number of features that make the code suitable for the study of ultra-relativistic jets.

The WENO scheme is one of upwind methods, designed to achieve a high-order accuracy in smooth regions and keep the essentially non-oscillatory property near discontinuities; hence, it can accurately reproduce the nonlinear dynamics in complex flows. The basic idea of the WENO scheme lies in an adaptive interpolation or reconstruction procedure \citep[see][for a review]{shu2009}. \citet{liu1994} first introduced a 3rd-order accurate finite volume (FV) WENO scheme, in which the cell averages integrated over the cell volume are used to obtain the reconstructed solutions at the cell interfaces. Later, \citet{jiang1996} proposed a 5th-order accurate finite difference (FD) WENO scheme, in which the point values of the physical fluxes are used to produce the reconstructed fluxes  at the cell interfaces with weight functions. Since the work of \citet{jiang1996}, different versions of FD WENO scheme with improved weight functions have been proposed to achieve higher resolution for smooth flows and/or smaller dissipation near discontinuities. Our code is built by implementing the three versions that have been widely applied to various computational fluid dynamic codes: (1) the WENO scheme presented in \citet{jiang1996} and \citet{jiang1999} (WENO-JS, hereafter), (2) WENO-Z by \citet{borges2008}, and (3) WENO-ZA by \citet{liu2018}, all of which have 5th-order spatial accuracy. A partial list of other FD WENO varieties includes WENO-M by \citet{henrick2005}, WENO-CU by \citet{hu2010}, and WENO-NS by \citet{ha2013}.

For the time integration of hyperbolic equations, the classical Runge-Kutta (RK) methods have been widely employed along with the WENO scheme \citep[e.g.,][]{shu1988,shu1989,jiang1996}. An improved approach, called the strong stability preserving Runge-Kutta (SSPRK) method, was proposed to ensure nonlinear stability properties and treat discontinuous structures without spurious oscillations and smearing \citep[e.g.,][]{spiteri2002,spiteri2003,gottlieb2005}. Since shocks and contact discontinuities are ubiquitous in the flows associated with relativistic jets, our code is built with the 4th-order accurate, 5-stage SSPRK time integration method described in \citet{spiteri2002}, as well as the 4th-order accurate Runge-Kutta (RK4) method for comparison.

For multi-dimensional problems, FD schemes can achieve high-order accuracies with the so-called dimension-by-dimension method, in which the one-dimensional (1D) flux is used along each direction \citep[e.g.,][]{shu1989}. On the other hand, FV schemes equipped with the dimension-by-dimension method normally retain only the 2nd-order accuracy in nonlinear problems \citep[e.g.,][]{zhang2011}. \citet{buchmuller2014} and \citet{buchmuller2016} proposed a modified dimension-by-dimension method for FV WENO schemes, which leads to high-order accuracies for smooth solutions. Although our code employs a FD WENO scheme, it borrows the idea of the modified dimension-by-dimension method and implements a high-order accurate averaging of fluxes along the transverse directions in smooth flow regions to improve the performance of multi-dimensional problems.

It is known that high-accurate, shock-capturing, upwind codes could be prone to the so-called carbuncle instability, resulting in the deformation of shock front \citep[e.g.,][]{peery1988,dumbser2004}. The instability develops in particular when a shock wave propagates almost aligned with the computation grid. It arises owing to the insufficient numerical dissipation of upwind schemes. WENO codes like ours are expected to be subject to this instability. In fact, in test simulations of relativistic jets with our code, the instability often appears at the bow shock surface that encompasses the cocoon and shocked ICM (see Section \ref{s3.3.2} below). ``Cures'' for this problem normally introduce additional numerical dissipations at shocks, but at the cost of losing accuracies of original schemes \citep[e.g.,][]{pandolfi2001,hanawa2008}. Recently, \citet{fleischmann2020} demonstrated that a modification of eigenvalues for the acoustic modes can suppress the carbuncle instability, while keeping the numerical dissipation low. Our code implements this modification to prevent this carbuncle phenomenon.

The description of RHD fluids requires, along with the governing equations for the dynamics, the equation of state (EOS) that relates the pressure to the internal energy density, that is, the thermodynamics. For the sake of convenience, we adopt the following short names for the EOSs used here: RP, ID, TM, and RC. RP stands for the EOS of the ``single-component''\footnote{Here, by ``single-component'', fluids are composed of particles of same mass. However, relativistic fluids may consist of electrons, positrons, protons, and ions, which become thermally relativistic at different temperatures. Hence, in general, whether a fluid is in thermally relativistic regime or not should depend not only on the temperature but also on its composition (i.e., either electron–positron pairs, or electrons and protons, or some other combinations). Our single-component EOSs below do not include the composition dependence of relativistic behaviors. An EOS for multi-component relativistic fluids is discussed, for instance, in \citet{chattopadhyay2009}, and its effects on shocks in relativistic jets are discussed, for instance, in \citet{josh2021}.} perfect gas in relativistic regime \citep[e.g.,][]{synge1957}. In RP, the relation between the pressure and the internal energy density is given with the modified Bessel functions, which is computationally expensive to be managed in numerical codes. Hence, the so-called ideal EOS (ID for short) assuming a constant adiabatic index $\gamma$  has been widely used in RHD codes. But with ID, the transition from $\gamma=5/3$ for subrelativistic temperature to $\gamma=4/3$ for fully relativistic temperature or vice versa can not be properly reproduced. On the other hand, EOSs that closely approximate RP and yet are simple enough to be used in numerical codes have been suggested. For instance, an EOS which follows the lower bound of the Taub's inequality \citep{taub1948} was first used by \citet{mathews1971} and later proposed for a RHD code by \citet{mignone2005b} (TM for short). Another EOS that gives a better fit to RP was suggested by \citet{ryu2006} (RC for short). Our code implements ID, TM and RC in order to examine the effects of different EOSs.

In this paper, we describe in detail the different elements of our newly developed RHD code, and present an extensive set of canonical tests, which are devised to demonstrate the accuracy and robustness of the code. We then choose the 5th-order accurate WENO-Z, the 4th-order accurate SSPRK, and RC as the fiducial setup for the simulation study of ultra-relativistic jets, which is reported in Paper II.

The paper is organized as follows. In Section \ref{s2} we present the RHD equations and EOSs, and describe the components of the code, such as the WENO schemes and the time integration methods along with additional features. The tests are presented in Section \ref{s3}. Summary and discussion follow in Section \ref{s4}.

\section{Code Description}
\label{s2}

\subsection{Basic Equations}
\label{s2.1}

The conservation equations that govern the special RHDs in the laboratory frame can be written as a hyperbolic system of equations
\begin{eqnarray}
\frac{\partial D}{\partial t} + \mbox{\boldmath$\nabla$}\cdot(D\mbox{\boldmath$v$}) = 0, \label{eq:De}\\
\frac{\partial \mbox{\boldmath$M$}}{\partial t} + \mbox{\boldmath$\nabla$}\cdot(\mbox{\boldmath$M$}\mbox{\boldmath$v$}+p) = 0, \label{eq:Me}\\
\frac{\partial E}{\partial t} + \mbox{\boldmath$\nabla$}\cdot[(E+p)\mbox{\boldmath$v$}] = 0, \label{eq:Ee}
\end{eqnarray}
where the conserved quantities, $D$, \mbox{\boldmath$M$}, and $E$, are the mass, momentum, and total energy densities, respectively \citep[e.g][]{landau1959}. The conserved quantities are related to the primitive variables, the rest-mass density, $\rho$, the fluid three-velocity, \mbox{\boldmath$v$}, and the isotropic gas pressure, $p$, as follows:
\begin{eqnarray}
D = \Gamma\rho, \label{eq:D}\\
\mbox{\boldmath$M$} = \Gamma^{2}\rho h \mbox{\boldmath$v$}, \label{eq:M}\\
E = \Gamma^{2}\rho h - p, \label{eq:E}
\end{eqnarray}
where $\Gamma=1/\sqrt{1-v^{2}}$ with $v=|\mbox{\boldmath$v$}|$ is the Lorentz factor, and $h=(e+p)/\rho$ is the specific enthalpy. Here, $e$ is the sum of the internal and rest-mass energy densities. The velocity is normalized by the speed of light (i.e., $c\equiv1$).

\subsection{Equation of State}
\label{s2.2}

The above equations for RHDs is ``closed'' with an EOS. In the RP EOS, the specific enthalpy is given as
\begin{equation}
h(p,\rho)=\frac{K_{3}(1/\Theta)}{K_{2}(1/\Theta)},
\label{eq:RP}
\end{equation}
where $K_{2}$ and $K_{3}$ are the modified Bessel functions of the second kind of order two and three, respectively \citep[e.g.,][]{synge1957}. Here, $\Theta=p/\rho$ is a temperature-like variable.

As mentioned in the introduction, RP is not a practical EOS for RHD codes, because the calculation of the primitive variables from the conserved quantities is computationally expensive due to the modified Bessel functions. Hence, most RHD codes have adopted either computationally manageable EOSs or EOSs that approximate RP, including the followings:\hfil\break
(1) the ID EOS with a constant adiabatic index $\gamma$,
\begin{equation}
p=(\gamma-1)(e-\rho)~~~~~{\rm or}~~~~~h=1+\frac{\gamma\Theta}{\gamma-1},
\end{equation}
(2) the TM EOS suggested by \citet{mignone2005b},
\begin{equation}
\frac{p}{\rho}=\frac{1}{3}\left(\frac{e}{\rho}-\frac{\rho}{e}\right)~~~~~{\rm or}~~~~~h=\frac{5}{2}\Theta+\frac{3}{2}\sqrt{\Theta^{2}+\frac{4}{9}},
\end{equation}
and (3) the RC EOS suggested by \citet{ryu2006},
\begin{equation}
\frac{p}{e-\rho}=\frac{3p+2\rho}{9p+3\rho}~~~~~{\rm or}~~~~~h=2\frac{6\Theta^{2}+4\Theta+1}{3\Theta+2}.
\end{equation}
With the relativistic kinetic theory, \citet{taub1948} showed that the EOS for RHDs should satisfy the inequality
\begin{equation}
(h-\Theta)(h-4\Theta)\ge1.
\end{equation}
While TM follows the lower bound of the inequality, RC locates above it satisfying the inequality for all $\Theta$.

The primitive variables, which are used in the calculation of numerical fluxes (see the next subsection), can be obtained through the inversion of Equations (\ref{eq:D}) - (\ref{eq:E}), following the procedures presented in \citet{ryu2006}. The three EOSs, ID, TM, and RC, are described in detail and compared to RP in \citet{ryu2006}. We point that while both RC and TM approximate RP, RC gives a slightly better fit to RP.

\subsection{Finite Difference WENO Scheme}
\label{s2.3}

In Cartesian geometry, Equations (\ref{eq:De}) - (\ref{eq:Ee}) can be written as
\begin{equation}
\frac{\partial\mbox{\boldmath$q$}}{\partial t}
+\frac{\partial\mbox{\boldmath$F$}}{\partial x}
+\frac{\partial\mbox{\boldmath$G$}}{\partial y}
+\frac{\partial\mbox{\boldmath$H$}}{\partial z} = 0 \label{eq:consEQ},
\end{equation}
with the state vector, \mbox{\boldmath$q$}, and the flux vectors, \mbox{\boldmath$F$}, \mbox{\boldmath$G$}, and \mbox{\boldmath$H$}. The explicit forms of the state and flux vectors can be found, for example, in \citet{ryu2006}. Our code updates the state vector, $\mbox{\boldmath$q$}_{i,j,k}$, defined at the cell center of three-dimensional (3D), uniform Cartesian grids, using the dimension-by-dimension method:
\begin{equation}
\mbox{\boldmath$q$}_{i,j,k}' = \mbox{\boldmath$q$}_{i,j,k}
- \frac{\Delta t}{\Delta x}\left(\mbox{\boldmath$F$}_{i+\frac{1}{2},j,k}-\mbox{\boldmath$F$}_{i-\frac{1}{2},j,k}\right)
- \frac{\Delta t}{\Delta y}\left(\mbox{\boldmath$G$}_{i,j+\frac{1}{2},k}-\mbox{\boldmath$G$}_{i,j-\frac{1}{2},k}\right)
- \frac{\Delta t}{\Delta z}\left(\mbox{\boldmath$H$}_{i,j,k+\frac{1}{2}}-\mbox{\boldmath$H$}_{i,j,k-\frac{1}{2}}\right),
\end{equation}
where $\Delta x$, $\Delta y$, and $\Delta z$ are the cell sizes along the $x$, $y$, and $z$-directions, and $\Delta t$ is the time step (see the next subsection). 
The numerical fluxes, $\mbox{\boldmath$F$}_{i\pm\frac{1}{2},j,k}$, $\mbox{\boldmath$G$}_{i,j\pm\frac{1}{2},k}$, and $\mbox{\boldmath$H$}_{i,j,k\pm\frac{1}{2}}$, defined at the cell interfaces, are estimated with the 5th-order accurate FD WENO scheme.

We here describe the reconstructions of the $x$-flux, $\mbox{\boldmath$F$}_{i\pm\frac{1}{2}}$ (the subscripts $j$ and $k$ are dropped for simplicity), while the reconstructions of $y$ and $z$-fluxes can be done by alternating the coordinates.
The numerical flux, $\mbox{\boldmath$F$}_{i+\frac{1}{2}}$, is obtained using the cell center fluxes, $\mbox{\boldmath$F$}_{m}$ at $m=i-2, \dots, i+3$, \citep{jiang1996,jiang1999} as
\begin{equation}
\begin{aligned}
\mbox{\boldmath$F$}_{i+\frac{1}{2}}=\frac{1}{12}\left(-\mbox{\boldmath$F$}_{i-1}+7\mbox{\boldmath$F$}_{i}+7\mbox{\boldmath$F$}_{i+1}-\mbox{\boldmath$F$}_{i+2}\right) ~~~~~~~~~~~~~~~~~~~~~~~~~~~~~~~~~~~~~ \\
+\sum_{s=1}^{5}\Big[- \mathbf{\varphi}_{N}\left(\Delta\mbox{\boldmath$F$}^{s+}_{i-\frac{3}{2}},\Delta\mbox{\boldmath$F$}^{s+}_{i-\frac{1}{2}},\Delta\mbox{\boldmath$F$}^{s+}_{i+\frac{1}{2}},\Delta \mbox{\boldmath$F$}^{s+}_{i+\frac{3}{2}}\right)
+ \mathbf{\varphi}_{N}\left(\Delta\mbox{\boldmath$F$}^{s-}_{i+\frac{5}{2}},\Delta\mbox{\boldmath$F$}^{s-}_{i+\frac{3}{2}},\Delta\mbox{\boldmath$F$}^{s-}_{i+\frac{1}{2}},\Delta \mbox{\boldmath$F$}^{s-}_{i-\frac{1}{2}}\right)\Big]\mbox{\boldmath$R$}^{s}_{i+\frac{1}{2}},
\end{aligned}
\end{equation}
where $s$ denotes the five characteristic modes of RHDs, and $\mbox{\boldmath$R$}^{s}_{i+\frac{1}{2}}$ is the right eigenvector. Here, $\Delta\mbox{\boldmath$F$}^{s\pm}_{m+\frac{1}{2}}$ with $m=i-2, \dots, i+2$ is calculated using the local Lax-Friedrichs flux splitting as follows:
\begin{eqnarray}
\mbox{\boldmath$F$}^{s}_{m}=\mbox{\boldmath$L$}^s_{i+\frac{1}{2}}\mbox{\boldmath$F$}_{m}, ~~~~~
\mbox{\boldmath$q$}^{s}_{m}=\mbox{\boldmath$L$}^s_{i+\frac{1}{2}}\mbox{\boldmath$q$}_{m}, ~~~~~~~~~~\\
\Delta\mbox{\boldmath$F$}^{s}_{m+\frac{1}{2}}=\mbox{\boldmath$F$}^{s}_{m+1}-\mbox{\boldmath$F$}^{s}_{m}, ~~
\Delta\mbox{\boldmath$q$}^{s}_{m+\frac{1}{2}}=\mbox{\boldmath$q$}^{s}_{m+1}-\mbox{\boldmath$q$}^{s}_{m}, ~~~\\
\Delta\mbox{\boldmath$F$}^{s\pm}_{m+\frac{1}{2}} = \frac{1}{2}\left(\Delta\mbox{\boldmath$F$}^{s}_{m+\frac{1}{2}}\pm\mathbf{\delta}_{i+\frac{1}{2}}^{s}\Delta\mbox{\boldmath$q$}^{s}_{m+\frac{1}{2}}\right), ~~~~~~~~ \label{eq:DFpm}
\end{eqnarray}
where $\mathbf{\delta}_{i+\frac{1}{2}}^{s}=\max_{i-2\leq m\leq i+3}|\lambda^s_m|$ with the $s$th eigenvalue $\lambda^s_m$, and $\mbox{\boldmath$L$}^{s}_{i+\frac{1}{2}}$ is the left eigenvector.

The above steps require the eigenvalues and eigenvectors of characteristic modes. We use the formulae presented in \citet{ryu2006}; they are expressed in generic forms with the primitive variables, specific enthalpy, polytropic index, and sound speed, and hence can be used independent of EOSs. The eigenvalues, $\lambda^s_i$, are calculated using the primitive variables at the cell center, while the eigenvectors, $\mbox{\boldmath$R$}^{s}_{i+\frac{1}{2}}$ and $\mbox{\boldmath$L$}^{s}_{i+\frac{1}{2}}$, at the cell interfaces are calculated using the arithmetic averagings of the fluid three-velocity and specific enthalpy at the two neighboring grids of $i$ and $i+1$.

The function $\mathbf{\varphi}_{N}$ is defined as
\begin{equation}
\mathbf{\varphi}_{N}(a,b,c,d) = \frac{1}{3}\omega_{0}(a-2b+c)+\frac{1}{6}\left(\omega_{2}-\frac{1}{2}\right)(b-2c+d).
\end{equation}
Here, $\omega_0$ and $\omega_2$ are the weight functions, which are designed to achieve high-order accuracies in smooth flows, while keeping the essentially non-oscillatory property near discontinuities. Since the widely used WENO-JS \citep{jiang1996}, different versions of WENO scheme with different weight functions have been suggested to improve performance, such as WENO-Z by \citet{borges2008} and WENO-ZA by \citet{liu2018}.

The weight functions are given as\footnote{Note that $\delta$'s here denote quantities different from $\delta$ in Equation (\ref{eq:DFpm}); $\delta$ is used in both Equations (\ref{eq:DFpm}) and (\ref{eq:alphaJS}) to keep the original notations of \citet{jiang1996}.}
\begin{equation}
\omega_{0}={\frac{\delta_{0}}{{\delta_{0}+\delta_{1}+\delta_{2}}}}, ~~~~ \omega_{2}={\frac{\delta_{2}}{{\delta_{0}+\delta_{1}+\delta_{2}}}}.
\end{equation}
In WENO-JS,
\begin{equation}
\delta^{JS}_{r}={\frac{C_{r}}{{(\epsilon+IS_{r})^{2}}}}, ~~~~ r=0,1,2, \label{eq:alphaJS}
\end{equation}
in WENO-Z,
\begin{equation}
\delta^{Z}_{r} = {C_{r}\left(1+\left(\frac{\tau_{5}}{{\epsilon+IS_{r}}}\right)^2\right)}, ~~~~ r=0,1,2,
\end{equation}
and in WENO-ZA,
\begin{equation}
\delta^{ZA}_{r} = {C_{r}\left(1+\frac{A\cdot\tau_6}{\epsilon+IS_{r}}\right)}, ~~~~ r=0,1,2,
\end{equation}
where $C_{0}=1$, $C_{1}=6$, and $C_{2}=3$, respectively. Here, $IS_{r}$'s are the local smoothness indicators, which are given as
\begin{equation}
\begin{aligned}
IS_{0} = 13(a-b)^{2}+3(a-3b)^{2},\\
IS_{1} = 13(b-c)^{2}+3(b+c)^{2},\\
IS_{2} = 13(c-d)^{2}+3(3c-d)^{2}.
\end{aligned}
\end{equation}
In WENO-Z, $\tau_{5}=\left|IS_0-IS_2\right|$, and in WENO-ZA,
\begin{eqnarray}
A = {\frac{\tau_{6}}{{\epsilon+IS_{0}+IS_{2}-\tau_{6}}}}, ~~~~~~~~~~~~~~~~~~~~~~~\\
\tau_{6} = 3\left(\left|a-3b\right|-\left|3c-d\right|\right)^2
+13\left(\left|a-b\right|-\left|c-d\right|\right)^2.~~~~
\end{eqnarray}
The parameter $\epsilon$ is included to avoid the zero denominator, and $\epsilon=10^{-6}$ is used in our code.

WENO-Z and WENO-ZA, which include high-order global smoothness indicators, were developed to achieve higher resolution and smaller dissipation than WENO-JS, but at the same time, they could be prone to numerical artifacts for a wider range of problems.

\subsection{Time Integration}
\label{s2.4}

Numerical codes based on the WENO scheme commonly employ high-order RK-type methods for the time integration. Our code implements the 4th-order accurate, 5-stage SSPRK method \citep[e.g.,][]{spiteri2002,spiteri2003,gottlieb2005}, as well as RK4, the classical 4th-order accurate RK method \citep[e.g.,][]{shu1988,shu1989,jiang1996}.

In RK4, the time-stepping from $\mbox{\boldmath$q$}^{n}_{i,j,k}$ to $\mbox{\boldmath$q$}^{n+1}_{i,j,k}$ proceeds with the following four stages:
\begin{equation}
\begin{aligned}
\mbox{\boldmath$q$}^{(0)}=\mbox{\boldmath$q$}^{n}, ~~~~~
\mbox{\boldmath$q$}^{(1)}=\mbox{\boldmath$q$}^{(0)}+\frac{\Delta t}{2}\mbox{\boldmath$\cal L$}^{(0)}, ~~~~~
\mbox{\boldmath$q$}^{(2)}=\mbox{\boldmath$q$}^{(0)}+\frac{\Delta t}{2}\mbox{\boldmath$\cal L$}^{(1)}, ~~~~~~~~~ \\
\mbox{\boldmath$q$}^{(3)}=\mbox{\boldmath$q$}^{(0)}+{\Delta t}\mbox{\boldmath$\cal L$}^{(2)}, ~~~~~
\mbox{\boldmath$q$}^{n+1}=\frac{1}{3}\left(-\mbox{\boldmath$q$}^{(0)}+\mbox{\boldmath$q$}^{(1)}+2\mbox{\boldmath$q$}^{(2)}+\mbox{\boldmath$q$}^{(3)}\right)+\frac{\Delta t}{6}\mbox{\boldmath$\cal L$}^{(3)}. \label{eq:RK4}
\end{aligned}
\end{equation}
In SSPRK, the time-stepping is given as
\begin{equation}
\mbox{\boldmath$q$}^{(0)}=\mbox{\boldmath$q$}^{n}, ~~~~~
\mbox{\boldmath$q$}^{(l)}=\sum_{m=0}^{l-1}(\chi_{lm} \mbox{\boldmath$q$}^{(m)}+\Delta t\beta_{lm} \mbox{\boldmath$\cal L$}^{(m)}),~~ l = 1,2,\cdots,5, ~~~~~
\mbox{\boldmath$q$}^{n+1}=\mbox{\boldmath$q$}^{(5)}. \label{eq:SSPRK}
\end{equation}
The coefficients $\chi_{lm}$ and $\beta_{lm}$ can be found, for instance, in \citet{spiteri2002}. Here
\begin{equation}
\mbox{\boldmath$\cal L$}^{(l)}_{i,j,k} = -\frac{\mbox{\boldmath$F$}^{(l)}_{i+\frac{1}{2},j,k}-\mbox{\boldmath$F$}^{(l)}_{i-\frac{1}{2},j,k}}{\Delta x}
-\frac{\mbox{\boldmath$G$}^{(l)}_{i,j+\frac{1}{2},k}-\mbox{\boldmath$G$}^{(l)}_{i,j-\frac{1}{2},k}}{\Delta y}
-\frac{\mbox{\boldmath$H$}^{(l)}_{i,j,k+\frac{1}{2}}-\mbox{\boldmath$H$}^{(l)}_{i,j,k-\frac{1}{2}}}{\Delta z},
\end{equation}
where $\mbox{\boldmath$\cal L$}^{(l)}_{i,j,k}$ is calculated with $\mbox{\boldmath$q$}^{(l)}_{i,j,k}$. In Equations (\ref{eq:RK4}) and (\ref{eq:SSPRK}), the subscripts $i$, $j$, and $k$ are dropped for simplicity.

The time step, $\Delta t$, should be restricted by the CFL (Courant-Friedrichs-Levy) condition for the numerical stability,
\begin{equation}
\Delta t = {\rm CFL}/\left[\frac{\lambda^{\rm max}_{x}}{\Delta x}+\frac{\lambda^{\rm max}_{y}}{\Delta y}+\frac{\lambda^{\rm max}_{z}}{\Delta z}\right],
\end{equation}
where $\lambda^{\rm max}$'s are the maxima of the eigenvalues at the cell center, $\lambda^s_{i,j,k}$, along the $x$, $y$, and $z$-directions, respectively. While normally ${\rm CFL} \le 1$ is required for RK methods, ${\rm CFL} > 1$ is allowed with SSPRK \citep[e.g.,][]{spiteri2002,spiteri2003}. In our code, CFL = 0.8 is used as the default value, unless otherwise specified.

\subsection{Averaging of Fluxes along Transverse Directions}
\label{s2.5}

As mentioned in the introduction, borrowing the idea of the modified dimension-by-dimension method proposed by \citet{buchmuller2014} and \citet{buchmuller2016} for FV WENO schemes, our code implements an averaging of fluxes along the transverse directions as follows. Here, the averaging of the $x$-flux is described, and those of the $y$ and $z$-fluxes can be done similarly.

For the calculation of the numerical fluxes described in Section \ref{s2.3}, instead of $\mbox{\boldmath$q$}_{i,j,k}$, the averaged state vector,
\begin{equation}
\mbox{\boldmath$\bar{q}$}_{i,j,k} = \frac{1}{\Delta y\Delta z}\int_{\sigma_{yz}}\mbox{\boldmath$q$}_{i,j,k}dydz + {\cal O}(\Delta^p),
\end{equation}
is used. Here, $\sigma_{yz}=[y_{i}-{\Delta y}/{2},y_{i}+{\Delta y}/{2}]\times[z_{i}-{\Delta z}/{2},z_{i}+{\Delta z}/{2}]$, and ${\cal O}(\Delta^p)$ denotes the order of the averaging. Then, in the time-integration step to update the state vectors, which is described in Section \ref{s2.4}, instead of $\mbox{\boldmath$F$}_{i\pm\frac{1}{2},j,k}$, the averaged flux,
\begin{equation}
\mbox{\boldmath$\bar{F}$}_{i\pm\frac{1}{2},j,k} = \frac{1}{\Delta y\Delta z}\int_{\sigma_{yz}}\mbox{\boldmath$F$}_{i\pm\frac{1}{2},j,k}dydz - {\cal O}(\Delta^p)
\end{equation}
is used.  To avoid possible problems of degrading shocks and contact discontinuities and generating spurious oscillations there, the above averagings are applied only to the regions of smooth flows \citep[see][for the detail]{buchmuller2016}.

Considering the accuracies of the WENO scheme and the SSPRK method in our code, the 4th-order accurate averaging with $p=4$ is employed, which is given as,
\begin{equation}
\mbox{\boldmath$\bar{q}$}_{i,j,k} = \mbox{\boldmath$q$}_{i,j,k}
-\frac{1}{24}\left(\mbox{\boldmath$q$}_{i,j-1,k}-2 \mbox{\boldmath$q$}_{i,j,k}+ \mbox{\boldmath$q$}_{i,j+1,k}\right)
-\frac{1}{24}\left(\mbox{\boldmath$q$}_{i,j,k-1}-2 \mbox{\boldmath$q$}_{i,j,k}+ \mbox{\boldmath$q$}_{i,j,k+1}\right),
\end{equation}
and
\begin{equation}
\mbox{\boldmath$\bar{F}$}_{i\pm\frac{1}{2},j,k} = \mbox{\boldmath$F$}_{i\pm\frac{1}{2},j,k}
+\frac{1}{24}\left(\mbox{\boldmath$F$}_{i\pm\frac{1}{2},j-1,k}-2 \mbox{\boldmath$F$}_{i\pm\frac{1}{2},j,k}+ \mbox{\boldmath$F$}_{i\pm\frac{1}{2},j+1,k}\right)
+\frac{1}{24}\left(\mbox{\boldmath$F$}_{i\pm\frac{1}{2},j,k-1}-2 \mbox{\boldmath$F$}_{i\pm\frac{1}{2},j,k}+ \mbox{\boldmath$F$}_{i\pm\frac{1}{2},j,k+1}\right).
\end{equation}
Higher-order accurate averagings can also be done \citep[see][for instance, for a 5th-order accurate averaging]{buchmuller2014,buchmuller2016}.

Hereafter, this modification of transverse fluxes is termed as the {\it transverse-flux averaging} for short. We expect that this improves the performance of multi-dimensional problems, as shown in Section \ref{s3.2.5} below.

\begin{figure*}[t] 
\vskip 0.2 cm
\centerline{\includegraphics[width=0.9\linewidth]{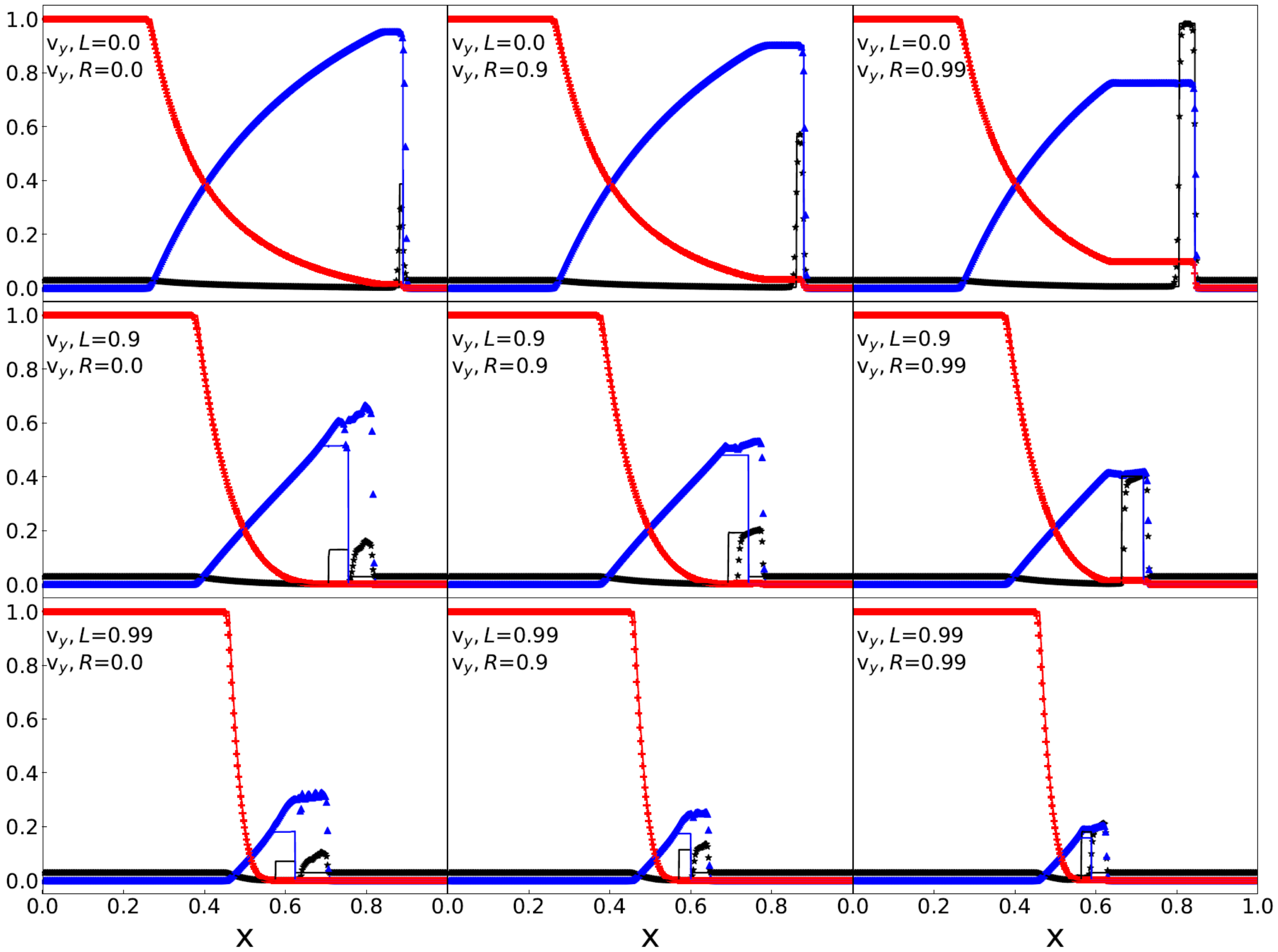}}
\vskip 0.0 cm
	\caption{1D relativistic shock tube tests with different combinations of transverse velocities. The initial condition is given in Equation (\ref{1dtest3}). The transverse velocities in the left ($x<0.5$) and right ($x>0.5$) sides are $v_{y,R}=(0,0.9,0.99)$ from left to right and $v_{y,L}=(0,0.9,0.99)$ from top to bottom, respectively. The results, $\rho/35$ (black asterisks), $v_x$ (blue triangles), and $p/10^3$ (red plus signs), of simulations with the default setup (WENO-Z, SSPRK, RC, and CFL = 0.8) using 400 grid zones are plotted at $t=0.4$. Simulations of very high-resolution ($2^{17}$ grid zones) are shown with the solid lines for comparison.}
\label{fig:f1}
\end{figure*}

\begin{figure*}[t] 
\vskip 0.2 cm
\hskip -0.1cm
\centerline{\includegraphics[width=0.5\linewidth]{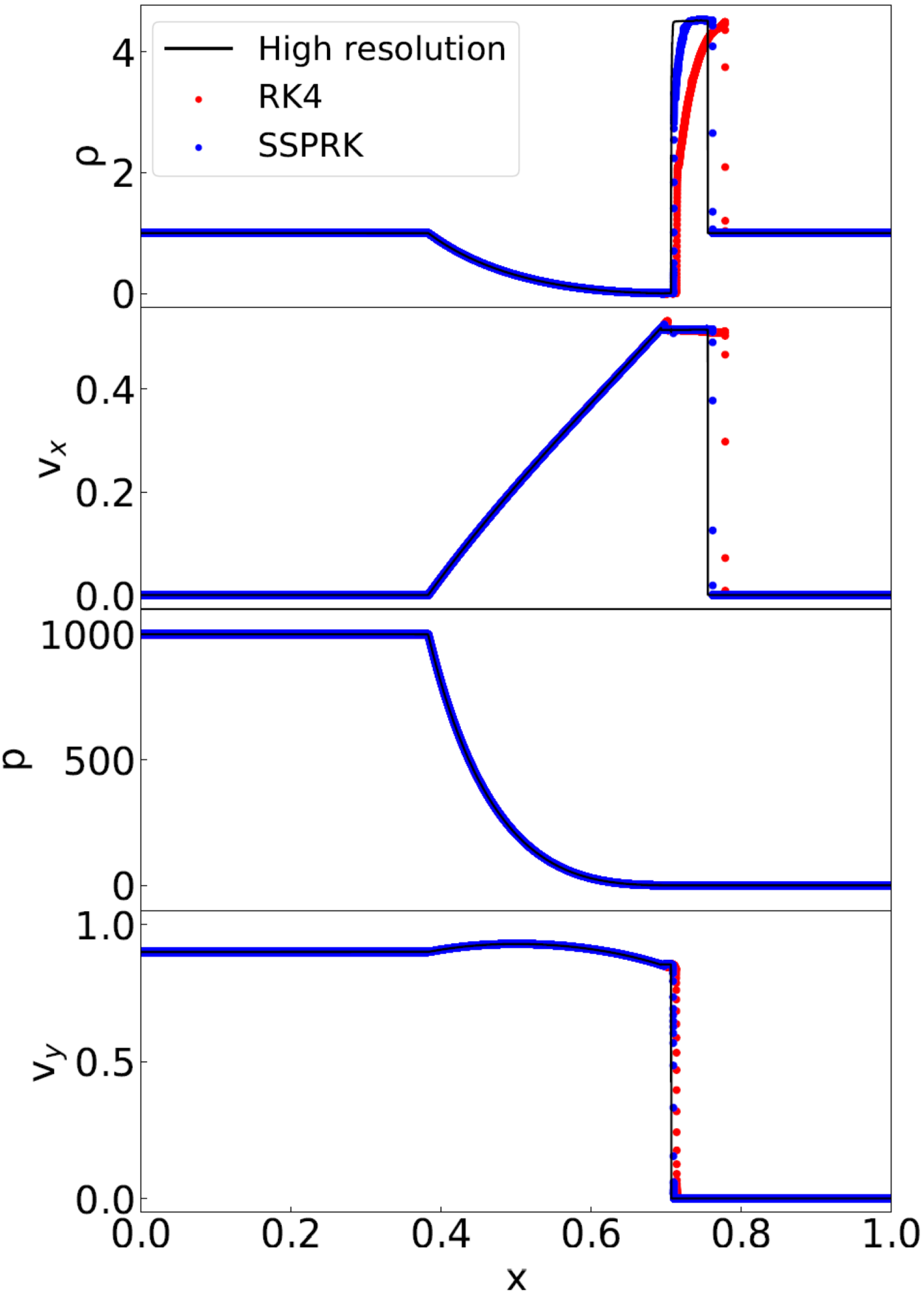}}
\vskip 0 cm
\caption{1D relativistic shock tube test to compare the RK4 (red) and SSPRK (blue) time integration methods. The initial condition is given in Equation (\ref{1dtest4}). The results of simulations with WENO-Z, RC, and CFL = 0.8 using 20,000 grid zones are plotted at $t = 0.45$. Simulations of very high-resolution ($2^{17}$ grid zones) are shown with the solid lines for comparison.}
\label{fig:f2}
\end{figure*}

\begin{figure*}[t] 
\vskip 0.2 cm
\hskip -0.1cm
\centerline{\includegraphics[width=0.5\linewidth]{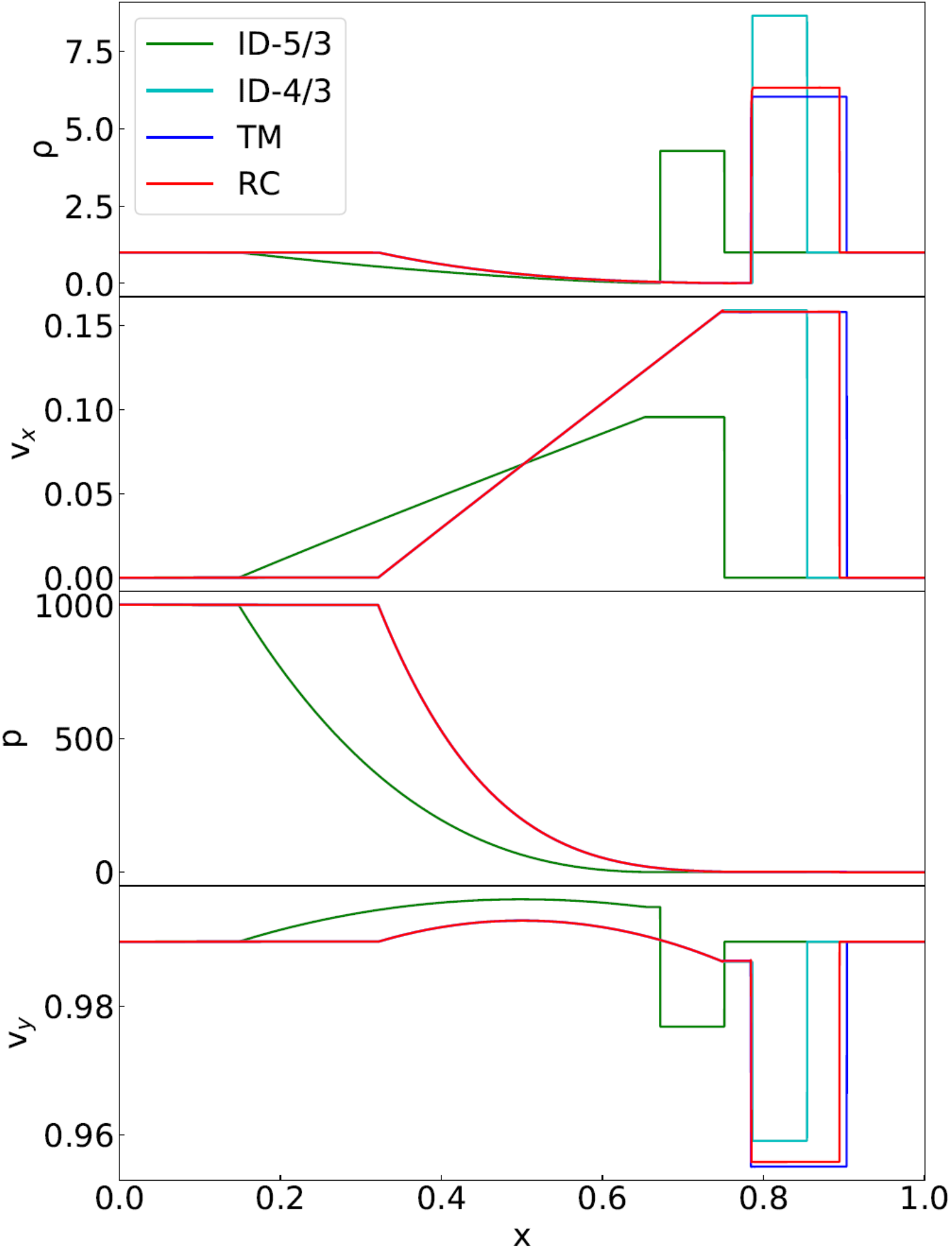}}
\vskip 0 cm
\caption{1D relativistic shock tube test to demonstrate the effects of different EOSs. The initial condition is given in Equation (\ref{1dtest1}). Simulations of very high-resolution ($2^{17}$ grid zones) adopting WENO-Z, SSPRK, and CFL = 0.8 are shown at $t = 1.8$.}
\label{fig:f3}
\end{figure*}

\begin{figure*}[t] 
\vskip 0.2 cm
\hskip -0.1cm
\centerline{\includegraphics[width=1\linewidth]{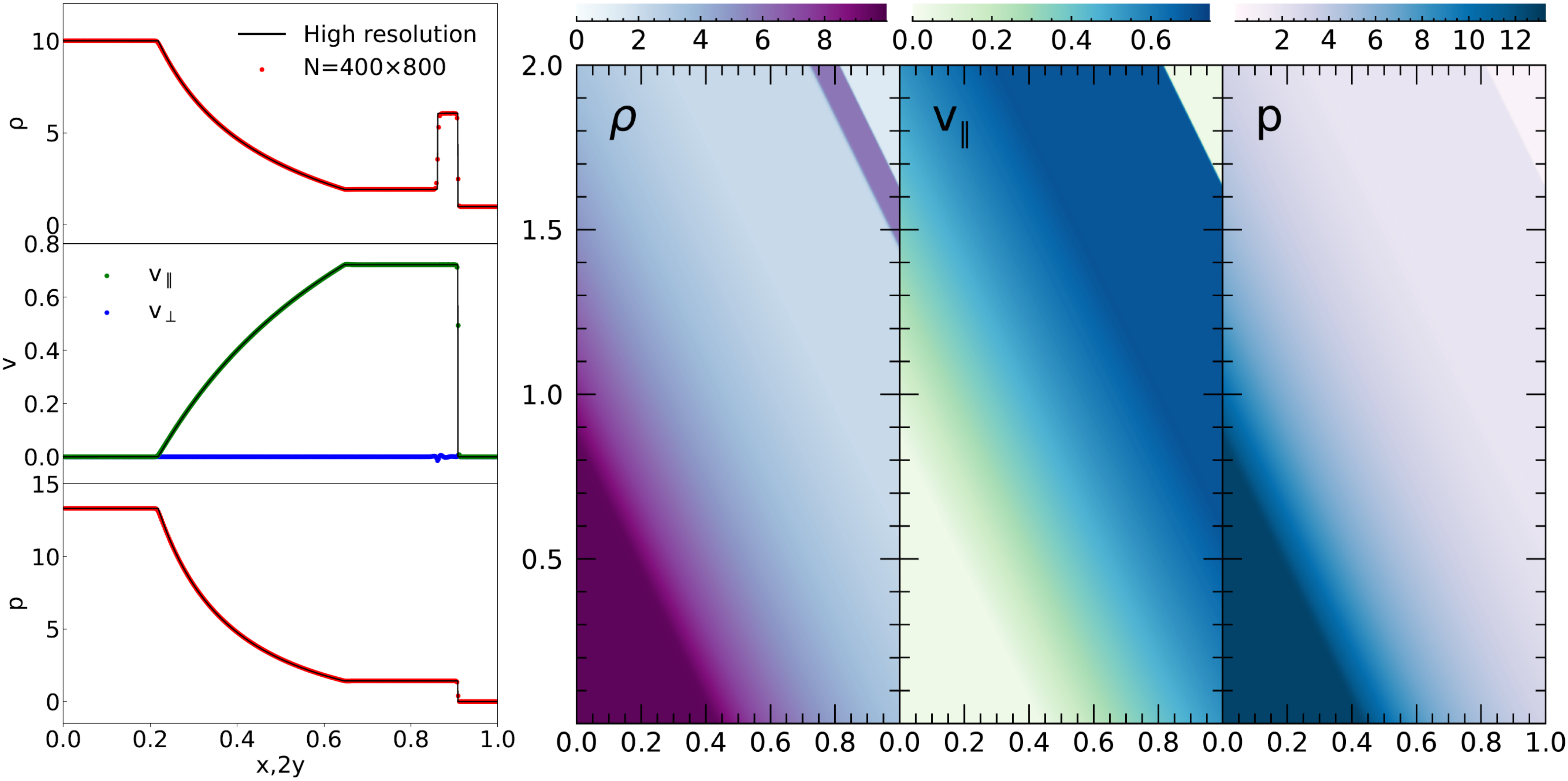}}
\vskip 0 cm
\caption{2D relativistic shock tube test. The initial condition is given in Equations (\ref{1dtest2-1}) and (\ref{1dtest2-2}). The results with the default setup (WENO-Z, SSPRK, RC, and CFL = 0.8) in the computational domain of [0,1]$\times$[0,2] using $400\times800$ grid zones are shown at $t = 0.4\times\sqrt{5}$. {\it Left panels}: The flow quantities along $x=y/2$ (dots) are compared to those from a 1D high-resolution simulation with 20,000 grid zones (black solid lines). In the middle panel, $v_{\parallel}$ (green dots) and $v_{\perp}$ (blue dots) are the fluid velocity components parallel and perpendicular to the propagation direction of structures. {\it Right panels}: The 2D images of the rest-mass density, $v_{\parallel}$, and pressure are shown.}
\label{fig:f4}
\end{figure*}

\subsection{Suppression of Carbuncle Instability}
\label{s2.6}

The carbuncle instability often appears at shocks in simulations based on upwind schemes including WENO, as mentioned in the introduction. If a shock propagates almost aligned with the computational grid in multi-dimensional problems, the disturbances traveling along the shock front at low Mach numbers tend to become unstable owing to insufficient dissipation. It is related to the well-known low Mach number problem of upwind schemes. \citet{fleischmann2020} suggested that the instability could be suppressed through a modification of eigenvalues for two acoustic modes as follows:
\begin{equation}
{c}'_{s} = \min(\phi|{v}_{x}|,{c}_{s}), ~~~~~ \mathbf{\lambda}_{1,5} = {{v}}_{x}\pm {c}'_{s},
\end{equation}
where $\phi$ is a positive number of order $\mathcal{O}(1)$. Then, the Mach numbers of the acoustic modes increase to $M\geq 1/\phi$ in low Mach number regions, and hence the modification in effect decreases the imbalance of advective and acoustic dissipations, easing the low Mach number problem. Following the suggestion of \citet{fleischmann2020}, we modify the eigenvalues of two acoustic modes in RHDs by limiting the local sound speed:
\begin{equation}
{c}'_{s} = \min(\phi|{v}_{x}|,{c}_{s}),~~~~~
\mathbf{\lambda}_{1,5} = \frac{(1-{c'}_{s}^{2})v_{x}\pm {c'}_{s}/\Gamma \sqrt{\mathcal{Q}} } {1-{c'}_{s}^{2}v^{2}},
\label{carb_cure}
\end{equation}
where $\mathcal{Q} = 1- v_x^2 - {c'}_s^2(v_y^2+v_z^2)$ (see Equation (43) of \citet{ryu2006}). 

As presented in Section \ref{s3.3.2} below, the carbuncle instability appears and deforms the bow shock surface in some of relativistic jet simulations with our code. It is demonstrated that the above modification effectively suppresses the instability, while the flow structures away from the bow shock are almost intact.

\section{Code Verification Tests}
\label{s3}

We have carried out a number of numerical tests to evaluate the performance of our RHD code described in the previous section. The default setup of the code consists of WENO-Z, SSPRK, and RC along with CFL = 0.8, and includes the 4th-order accurate transverse-flux averaging for multi-dimensional problems, unless stated otherwise.

\subsection{1D Tests: Relativistic Shock Tubes}
\label{s3.1}

Relativistic shock tube tests are set up with different RHD states in the left and right regions of the computational domain of $x=[0,1]$ with the initial discontinuity at $x=0.5$. In Newtonian hydrodynamics, the transverse velocities, that is, $y$ and $z$-velocities, do not affect the solutions of shock tube problems. In relativistic regime, on the other hand, the Lorentz factor depends on the transverse velocities, so do the solutions. Hence, in papers for RHD codes, tests have normally included nonzero transverse velocities (see the references in the introduction). Below, we present shock tube tests, setting the transverse velocity along the $y$-direction without loss of generality; then, the primitive variables of a RHD state is given as $\mbox{\boldmath$u$} = (\rho, v_{x},v_{y}, p)$, and the left and right states are denoted by $\mbox{\boldmath$u$}_{L}$ and $\mbox{\boldmath$u$}_{R}$.

We point that with nonzero $v_y$, the tests implicitly assume $v_y$ in $-\infty<y<\infty$ and the simultaneous occurrence of events along the $y$-axis at $t=0$, potentially violating the causality in the fluid-rest frame. While such setup is unphysical, that is the limitation of 1D tests. In addition, shocks with nonzero transverse velocity commonly appear in relativistic jets, and a clear example is the recollimation shock that forms in the jet spine flow (see Paper II). So we have also performed shock tube tests with nonzero $v_y$ and show them below.

The first test comprises shock tube problems with different combinations of transverse velocities in the left and right states, where the initial condition is given as
\begin{equation}
\mbox{\boldmath$u$}_{L} =(1,0,v_{y,L},10^{3}), ~~ \mbox{\boldmath$u$}_{R} =(1,0,v_{y,R},10^{-2}),
\label{1dtest3}
\end{equation}
with $v_{y,L}=0$, 0.9, 0.99 and $v_{y,R}=0$, 0.9, 0.99. Here, $\rho$, $v_x$ and $p$ are the same as in Equation (\ref{1dtest1}), and hence, the case with $v_{y,L}=0.99$ and $v_{y,R}=0.99$ is the same as in Figure \ref{fig:f3}. \citet{pons2000} presented the analytic solutions for these shock tube problems in the case of the ID EOS with $\gamma=5/3$. \citet{mignone2005b} employed these as a test for their PPM-based RHD code with the ID EOS.

Following \citet{mignone2005b}, Figure \ref{fig:f1} shows the results with RC using 400 grid zones for nine combinations of $v_{y,L}$ and $v_{y,R}$. Since the exact analytic solution is not available for the RC EOS, we compare the results with those of high-resolution simulations. Apart from the overall differences due to the different EOS employed, the trend in the numerical solutions of our WENO code is similar to those of the PPM code (see the Figure 5 of \citet{mignone2005b}); the solutions degrade when the transverse velocity is included. It turns out to be a very ``severe'' test, especially when the transverse velocity is large with $v_{y,L} \geq v_{y,R}$; the contact discontinuity shows further smearing and the shock location deviates from the correct position. While \citet{mignone2005b} attributed these features possibly to relativistic effects, we find that such degrading occurs with relativistic transverse velocities, independent of the EOS and numerical schemes. The overall capability of our WENO code to capture shocks and contact discontinuities in this test looks comparable to that of the PPM code.

The second test intends to compare the RK4 and SSPRK time integration methods. The initial condition is given as
\begin{equation}
\mbox{\boldmath$u$}_{L} =(1,0,0.9,10^{3}), ~~ \mbox{\boldmath$u$}_{R} =(1,0,0,10^{-2}).
\label{1dtest4}
\end{equation}
This is the same problem as that in the middle-left panel of Figure \ref{fig:f1}, and is one of the cases where the reproduction of acceptable numerical solutions requires a large number of grid zones. Figure \ref{fig:f2} shows the results using 20,000 grid zones. This test illustrates that SSPRK performs better than RK4, in terms of reproducing the correct solutions with a limited number of grid zones. Hence, we adopt SSPRK as the default time integration method. Note that with SSPRK, although the computation time is about $25~\%$ longer per time step, a larger CFL number is allowed; hence, the overall computational efficiency could be comparable in SSPRK and RK4.

We next examine the dependence of the results of a shock tube test on different EOSs. The initial condition is given as
\begin{equation}
\mbox{\boldmath$u$}_{L} =(1,0,0.99,10^{3}), ~~ \mbox{\boldmath$u$}_{R} =(1,0,0.99,10^{-2}).
\label{1dtest1}
\end{equation}
Figure \ref{fig:f3} shows the results of very high-resolution simulations with $2^{17}$ grid zones, demonstrating the differences due to different EOSs. This is the same as that in the bottom-right panel of Figure \ref{fig:f1}. This test was also presented in \citet{ryu2006}, and the results agree with those in their Figure 6 (which were produced with a different code). The numerical solutions with ID, both of $\gamma=5/3$ and $4/3$, are quite different from those with RC and TM. Considering that RC and TM are designed to fit the EOS for perfect relativistic gas, RP, it is clear that the simulations with ID fail to reproduce the correct solutions of RHDs with the RP EOS. The differences in the RC and TM solutions are relatively small. Yet, as mention in Section \ref{s2.2}, RC approximates RP better than TM, so we expect that the simulation using RC reproduces the solution with RP more closely. Hence, we here adopt RC as the fiducial EOS for our code.

\subsection{2D Tests}
\label{s3.2}

\subsubsection{2D Relativistic Shock Tube}
\label{s3.2.1}

The first two-dimensional (2D) test is a shock tube problem where the shock, contact discontinuity and rarefaction wave propagate along the off-axis in a rectangular domain. It is designed to evaluate the ability of our code to capture discontinuous structures and reproduce waves in 2D. The initial condition is given as
\begin{equation}
\mbox{\boldmath$u$}(x,y)=\begin{cases} \mbox{\boldmath$u$}_{L} & \text{ for $y<2(1-x)$,} \\
\mbox{\boldmath$u$}_{R} & \text{ for $y>2(1-x)$,} \end{cases}
\label{1dtest2-1}
\end{equation}
where
\begin{equation}
\mbox{\boldmath$u$}_{L} =(10,0,0,13.3), ~~ \mbox{\boldmath$u$}_{R} =(1,0,0,10^{-6}).
\label{1dtest2-2}
\end{equation}
Here and below, $\mbox{\boldmath$u$} = (\rho, v_{x},v_{y}, p)$ in 2D tests. The computational domain consists of [0,1]$\times$[0,2]. Here, the surfaces of the shock and contact discontinuity are formed with the inclination angle of $\arctan(1/2)=26.565^{\circ}$ with respect to the $x$-axis, and the continuous boundary inclined with the same angle is applied (see the right panels of Figure \ref{fig:f4}). This is the 2D version of a popular 1D shock tube test which has been widely performed in previous works \citep[e.g.,][and the papers that cite it]{schneider1993}.

Figure \ref{fig:f4} shows the results using $400\times800$ grid zones. Since the exact analytic solution is not available for the RC EOS, we again compare the results with those of a 1D high-resolution simulation. The smooth structure of rarefaction wave is well reproduced, while the shock and contact discontinuity are resolved with $2-4$ grid zones. In the left-middle panel, along with the fluid velocity parallel to the propagation direction of the structures, $v_{\parallel}$ (green dots), the transverse velocity, $v_{\perp}$ (blues dots), are shown. While $v_{\perp}=0$ expected, small, non-zero $v_{\perp}$ appears and it may be used to estimate the numerical error involved in capturing the shock and contact discontinuity in this test. Compared to $v_{\parallel}$ behind the shock, which is 0.721, we find that $|v_{\perp}|$ at the shock and contact discontinuity is at most $\sim1-2\ \%$.

\subsubsection{Convergence of Code}
\label{s3.2.2}

\begin{deluxetable}{ccccc}[t]
\tablecaption{2D Advection Test \label{tab:t1}}
\tablenum{1}
\tabletypesize{\small}
\tablecolumns{4}
\tablewidth{0pt}
\tablehead{
\colhead{$N\times2N^a$} &
\colhead{ } &
\colhead{${\Delta_N}^b$} &
\colhead{ } &
\colhead{${R_N}^b$}
}
\startdata
$20\times40$   && $2.69\times10^{-4}$ && - \\
$40\times80$   && $1.66\times10^{-5}$ && 4.02 \\
$80\times160$  && $1.03\times10^{-6}$ && 4.01 \\
$160\times320$ && $6.44\times10^{-8}$ && 4.00 \\
\enddata
\tablenotetext{a}{Resolution}
\tablenotetext{b}{$L_2$ error and the order of convergence}
\end{deluxetable}

We next consider a 2D advection test to quantify the accuracy of our code in the default setup. The initial condition is given as
\begin{equation}
\rho(x,y) = 1 + 0.2 \sin\left[2\pi(x\cos\theta+y\sin\theta)\right], ~~~
v_{x}(x,y) = 0.2,~~~v_{y}(x,y) = -0.1,~~~p(x,y)=1,
\end{equation}
with $\theta = 30^{\circ}$ in the periodic $x-y$ domain of $[0,2/\sqrt{3}]\times[0,2]$, represented with $N\times2N$ grid zones. Here, the density fluctuation propagates with the angle of $\arctan(-1/2)=-26.565^{\circ}$ with respect to the $x$-axis. We note that this is a test similar to those performed before, for instance, in \citet{he2012} and \citet{nunez2016}. While those previous studies set $v_{y}=0$, our test includes nonzero $v_{y}$ to make it a true 2D test.

The $L_2$ error and the order of convergence are estimated with the rest-mass density as
\begin{eqnarray}
\Delta_N = \frac{\sqrt{\sum^{N}_{i=1}\sum^{2N}_{j=1}(\rho^{\rm num}_{i,j}-\rho^{\rm exact}_{i,j})^{2}}}{\sqrt{\sum^{N}_{i=1}\sum^{2N}_{j=1}(\rho^{\rm exact}_{i,j})^{2}}},\\
R_N = \log_{2}\left(\frac{\Delta_{N/2}}{\Delta_N}\right).~~~~~~~~~~~
\end{eqnarray}
Here, $\rho^{\rm num}_{i,j}$ is the numerical solution, while $\rho^{\rm exact}_{i,j}$ is the exact solution. Table \ref{tab:t1} gives $\Delta_N$ and $R_N$ at $t=1$ for different numerical resolutions. $R_N\approx 4.0$ implies that our code achieves the 4th-order convergence, possibly due to the 4th-order accuracy of the SSPRK time integration method.

\subsubsection{2D Riemann Problem}
\label{s3.2.3}

One of conventional 2D tests is the Riemann problem, which consists of four square domains with different initial states \citep[e.g.,][]{del2002,mignone2005b,he2012,nunez2016}. This test is designed to evaluate how a multi-dimensional code handles the evolutions of shocks, contact discontinuities, and a jet-like structure. We consider, for instance, the specific problem performed by \citet{he2012} with the following initial condition:
\begin{equation}
\begin{cases} 
\mbox{\boldmath$u$}_{1} =(0.035145216124503,0,0,0.162931056509027), \quad x>0, y > 0, \\
\mbox{\boldmath$u$}_{2} =(0.1,0.7,0,1), \quad \quad \quad \quad \quad \quad \quad \quad \quad \quad \quad \quad \quad \quad \quad x<0,y>0, \\
\mbox{\boldmath$u$}_{3} =(0.5,0,0,1), \quad \quad \quad \quad \quad \quad \quad \quad \quad \quad \quad \quad \quad \quad \quad ~~ x<0,y<0, \\
\mbox{\boldmath$u$}_{4} =(0.1,0,0.7,1), \quad \quad \quad \quad \quad \quad \quad \quad \quad \quad \quad \quad \quad \quad \quad x>0,y<0, \end{cases}
\label{2DRiemann}
\end{equation}
in the computational domain of [-1,1]$\times$[-1,1], covered with $600\times600$ grid zones. The continuous boundary is imposed on the four faces.

\begin{figure*}[t] 
\vskip 0.2 cm
\centerline{\includegraphics[width=0.7\linewidth]{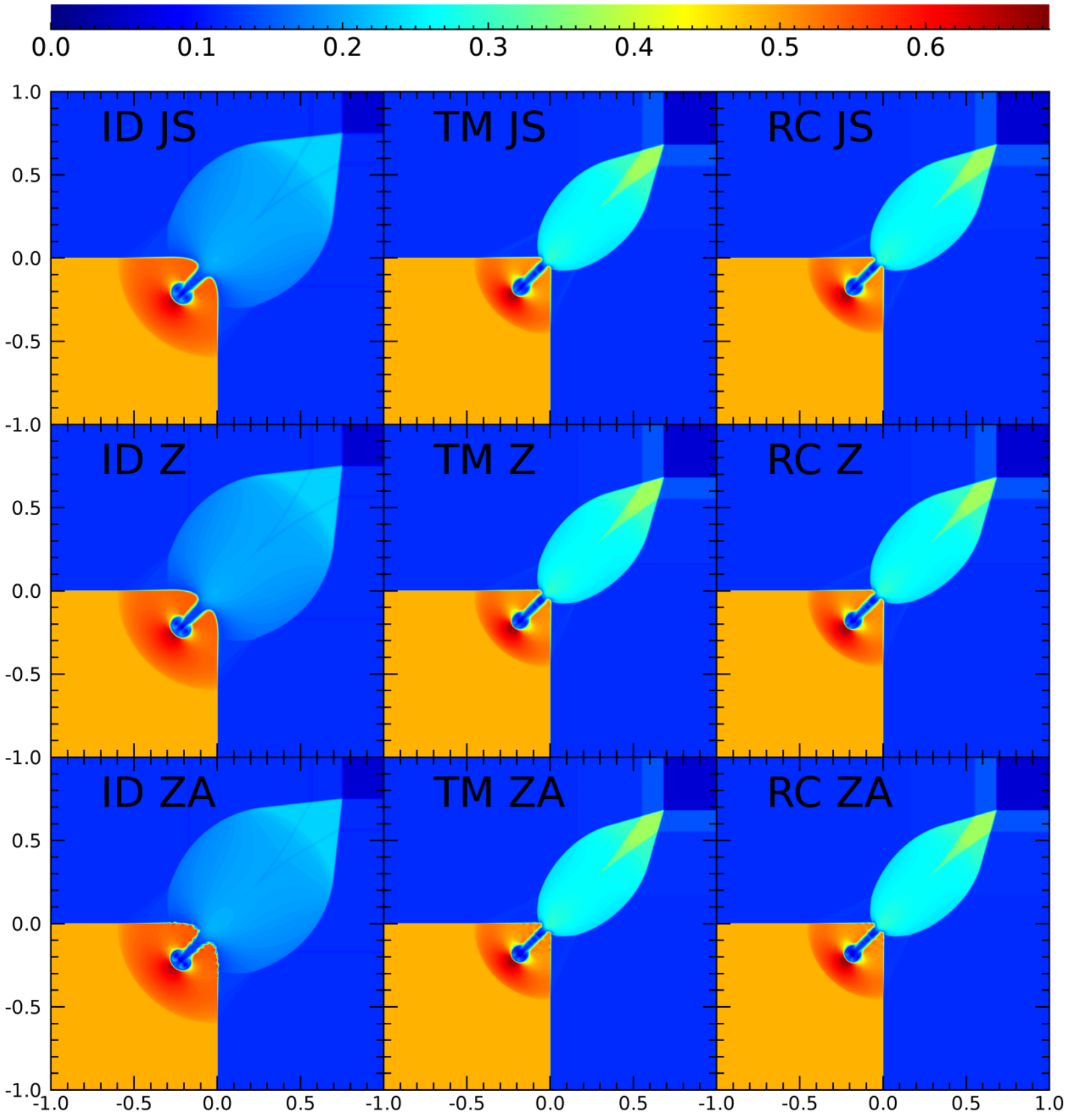}}
\vskip 0. cm
\caption{2D Riemann problem with the initial condition given in Equation (\ref{2DRiemann}). The rest-mass density, $\rho$, from simulations with different EOSs, ID with $\gamma=5/3$ (left), TM (center), and RC (right), using different WENO versions, WENO-JS (top), WENO-Z (middle), and WENO-ZA (bottom), is shown at $t = 0.8$. In all the simulations, the transverse-flux averaging is implemented, and the SSPRK time integration method with CFL = 0.8 is used. The computational domain of $[-1,1]\times[-1,1]$ is represented with $600\times600$ grid zones.}
\label{fig:f5}
\end{figure*}

Simulations have been carried out for nine combinations of different EOSs and different WENO versions, and are compared in Figure \ref{fig:f5}. The case of the ID EOS with $\gamma=5/3$ reproduces the results of previous studies \citep[see, e.g., Figure 5 of][]{he2012}. However, the ID results are clearly different from those with the RC and TM EOSs. Hence, this should be a multi-dimensional example, which illustrates the differences induced by the use of different EOSs. 
Looking closely at the center region of the domain, the simulations with WENO-ZA (bottom panels) display wiggles in the jet-like structure in blue, which expands in the lower-left direction. By contrast, those features are less noticeable in the simulations with WENO-JS and WENO-Z.

\begin{figure*}[t] 
\vskip 0.2 cm
\centerline{\includegraphics[width=0.7\linewidth]{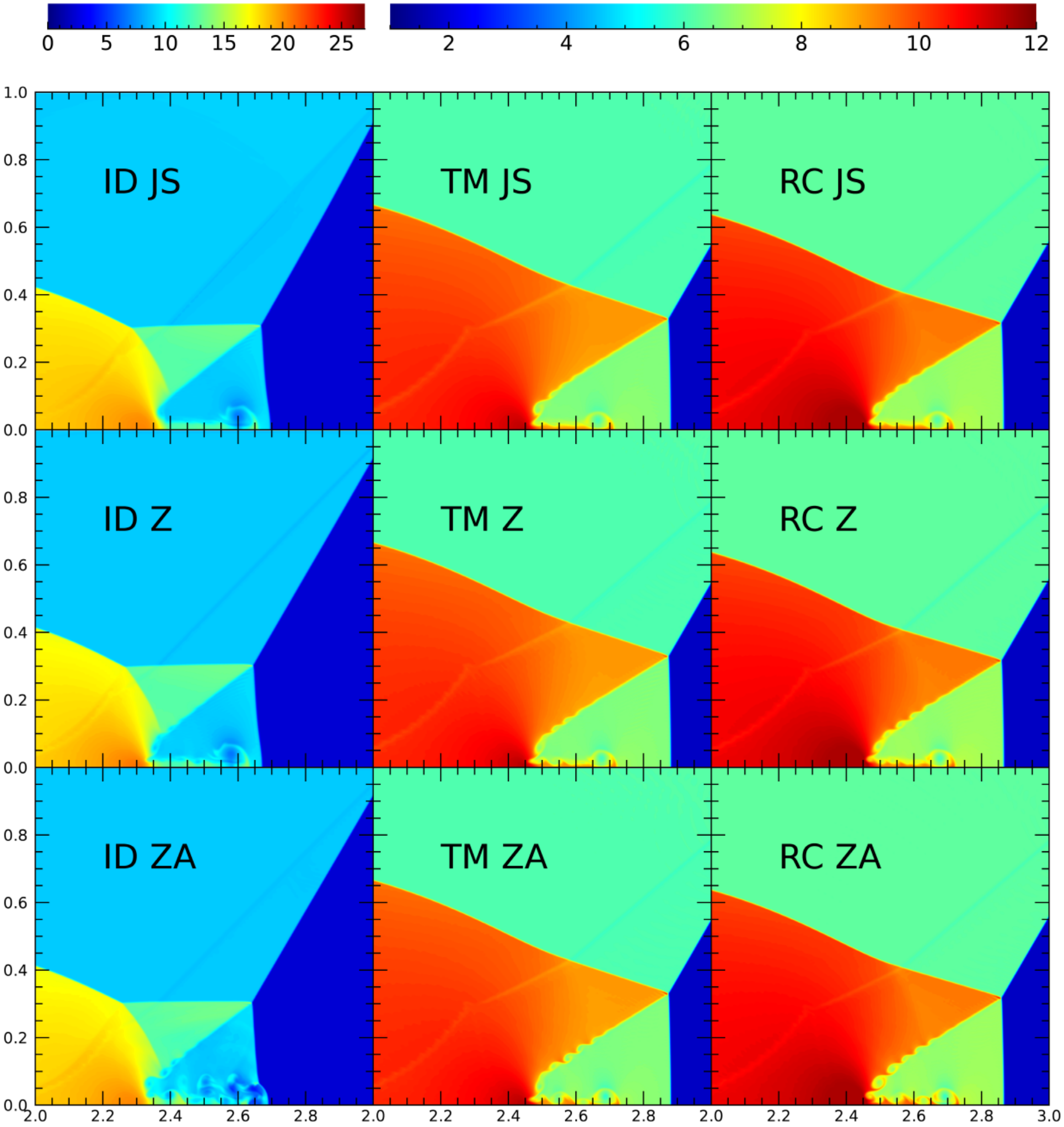}}
\vskip 0. cm
\caption{Relativistic double-Mach reflection problem with an inclined shock of $M_s=10$. The rest-mass density, $\rho$, from simulations with different EOSs, ID with $\gamma=1.4$ (left), TM (center), and RC (right), using different WENO versions, WENO-JS (top), WENO-Z (middle), and WENO-ZA (bottom), is shown at $t = 4$. The initial conditions are given in Equation (\ref{DM1}) for ID, in Equation (\ref{DM2}) for TM, and in Equation (\ref{DM3}) for RC. While the computational domain includes $[0,4]\times[0,1]$ with $1600\times400$ grid zones, only the region of $[2,3]\times[0,1]$ with $400\times400$ grid zones is plotted. The transverse-flux averaging is implemented, and the SSPRK time integration method with CFL = 0.8 is used, except in the case of ID and WENO-ZA (bottom left panel) where CFL = 0.5 is used. Note that the color bar for the ID case is different from that for the TM and RC cases.}
\label{fig:f6}
\end{figure*}

\begin{figure*}[t]
\vskip 0.2 cm
\hskip -0.1 cm
\centerline{\includegraphics[width=0.9\linewidth]{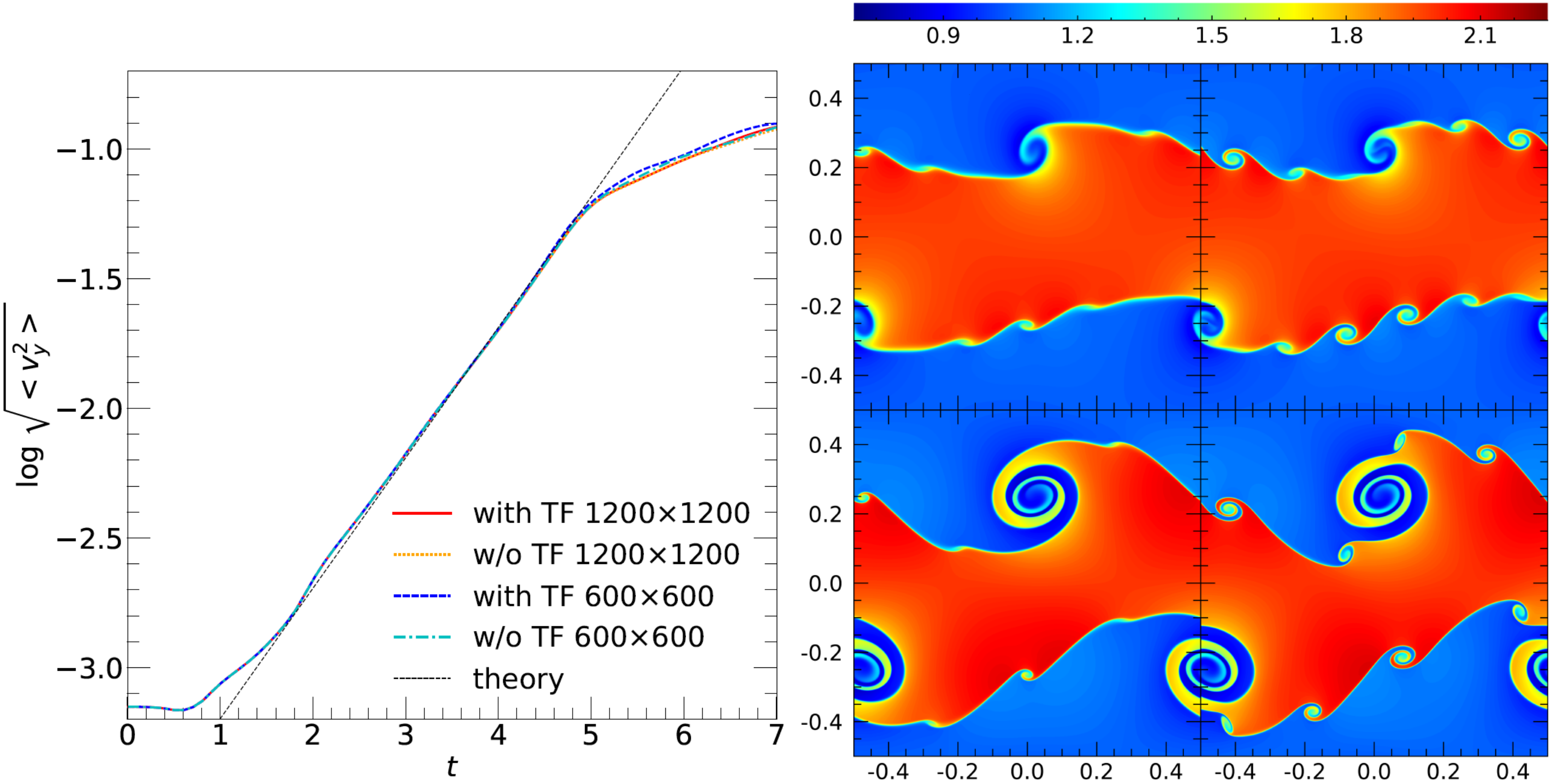}}
\vskip 0.0 cm
\caption{Relativistic KH instability test with the initial condition in Equation (\ref{KH1}). {\it Left panel}: The time evolution of the rms $y$-velocity, $\left<v_y^2\right>^{1/2}$, is plotted for simulations with/without transverse-flux averaging and with $600\times600$ and $1200\times1200$ grid zones. The black doted line draws the predicted linear growth of the instability from Equation (\ref{linearKH}). {\it Right panel}: The rest-mass density, $\rho$, is shown for simulations without (left) and with (right) transverse-flux averaging and with $600\times600$ grid zones at $t = 5$ (top) and $1200\times1200$ grid zones at $t = 6.5$ (bottom). The computational domain covers $[-0.5,0.5]\times[-0.5,0.5]$. The code equipped with the default setup (WENO-Z, SSPRK, RC, and CFL = 0.8) is used.}
\label{fig:f7}
\end{figure*}

\subsubsection{Double-Mach Reflection Problem}
\label{s3.2.4}

The double-Mach reflection test was first introduced by \citet{wood1984} for a Newtonian hydrodynamic code, and later employed for RHD codes in several studies \citep[e.g.,][]{ zhang2006, he2012,radice2012,nunez2016}. In this test, a shock of Mach number $M_{s}=10$ travels with the angle of $60^{\circ}$ with respect to the $x$-axis and is reflected at the wall along the $x$-axis, mimicking a shock colliding with an inclined wedge.

While previously the test was performed with the ID EOS ($\gamma=1.4$), we here consider the RC and TM EOSs as well. The initial condition for ID with $\gamma=4/3$ is given as
\begin{equation}
\mbox{\boldmath$u$}(x,y)=\begin{cases} \mbox{\boldmath$u$}_{L} & \text{ for $y>\sqrt{3}(x-1/6)$,} \\
\mbox{\boldmath$u$}_{R} & \text{ for $y<\sqrt{3}(x-1/6)$,} \end{cases}
\end{equation}
where 
\begin{equation}
\mbox{\boldmath$u$}_{L}= (8.564, 0.4247 \sin 60^\circ, -0.4247 \cos 60^\circ, 0.3808), ~~~~~ \mbox{\boldmath$u$}_{R}= (1.4, 0, 0, 0.0025),
\label{DM1}
\end{equation}
in the computational domain of $[0,4]\times[0,1]$ \citep[e.g.,][]{he2012}. The boundary condition is reflecting at the wall along the $x$-axis in $x>1/6$, and continuous elsewhere; at the top boundary of $y=1$, the condition is set to either $\mbox{\boldmath$u$}_{L}$ or $\mbox{\boldmath$u$}_{R}$ depending on the exact position of the shock.

The characteristics of the induced shock depend rather sensitively on the EOS. To mitigate the differences, we prescribe different postshock states $\mbox{\boldmath$u$}_{L}$ for different EOSs, while keeping the same preshock state $\mbox{\boldmath$u$}_{R}$ and $M_s=10$. Then, the initial condition is given as
\begin{equation}
\mbox{\boldmath$u$}_{L}= (6.0638, 0.4342 \sin 60^\circ, -0.4342 \cos 60^\circ, 0.4275)
\label{DM2}
\end{equation}
for the TM EOS, and
\begin{equation}
\mbox{\boldmath$u$}_{L}= (6.2506, 0.4340 \sin 60^\circ, -0.4340 \cos 60^\circ, 0.4322)
\label{DM3}
\end{equation}
for the RC EOS. Note that even with the same $\rho$ and $p$ in the preshock region, the sound speed, $c_s$, differs if the EOS is different (see, e.g., \citet{ryu2006} for the formulae of $c_s$ as a function of $p/\rho$ for different EOSs). So the shock speed, $v_s$, is different, even if the Mach number is fixed; in this test with $M_s=10$, $v_s=0.4984$, 0.5440, and 0.5438 for ID, TM, and RC, respectively. Again, this test demonstrates the differences owing to different EOSs.

Simulations have been performed with different WENO versions as well as with different EOSs, using $1600\times400$ grid zones, and compared in Figure \ref{fig:f6}. In this test, a contact discontinuity extends from a triple shock point to the lower-left direction, and is pushed by the high density and pressure blob and forms a jet-like structure along the $x$-axis. Figure \ref{fig:f6} focuses on the region of $[2,3]\times[0,1]$ that includes those structures. The contact discontinuity and jet-like structure are Kelvin–Helmholtz (KH) unstable, as was shown for the Newtonian test \citep[see, e.g.,][]{reyes2019}. In simulations with WENO-ZA, the KH vortices are well visible along the contact discontinuity, while they are less obvious with WENO-Z and almost absent with WENO-JS. In general, WENO-Z and WENO-ZA, equipped with higher-order global smoothness indicators, tend to produce richer structures than WENO-JS. On the other hand, in the case of ID and WENO-ZA (bottom left panel), the jet-like structure near the bottom reflecting wall looks distorted; as a matter of fact, in this case, the simulation crashes with CFL = 0.8, and hence CFL = 0.5 is used. This indicates that WENO-ZA may be more prone to numerical artifacts than WENO-JS and WENO-Z.

\subsubsection{Kelvin-Helmholtz Instability}
\label{s3.2.5}

The KH instability occurs at the interface of two slipping fluids, such as the jet-backflow interface in relativistic jets (see Paper II). It has been widely employed to evaluate the ability to handle the linear growth and the development of nonlinear structures in hydrodynamic codes \citep[e.g.,][for the test of RHD codes]{zhang2006,radice2012}. We here present a test of KH instability to illustrate the effects of the transverse-flux averaging, described in Section \ref{s2.5}, as well as the ability to reproduce the linear growth. The initial condition is given as
\begin{equation}
\mbox{\boldmath$u$}(x,y)=\begin{cases}( -(\rho_1-\rho_2)/2\tanh\left[(y-0.25)/\delta\right]+(\rho_1+\rho_2)/2,
                                        -U\tanh\left[(y-0.25)/\delta\right],0.001\sin(kx),p) & \text{for $y>0$,} \\
                                      (~~(\rho_1-\rho_2)/2\tanh\left[(y+0.25)/\delta\right]+(\rho_1+\rho_2)/2,
                                       ~~U\tanh\left[(y+0.25)/\delta\right],0.001\sin(kx),p) & \text{for $y<0$,} \end{cases}
\label{KH1}
\end{equation}
in the periodic domain of $[-0.5,0.5]\times[-0.5,0.5]$. Then, $\rho\approx\rho_1$ and $v_x\approx U$ at $y=0$, and $\rho\approx\rho_2$ and $v_x\approx-U$ at $y=\pm0.5$, with shear interfaces of thickness $\delta$ at $y=\pm0.25$.
We set $\rho_1=2$, $\rho_2=1$, $U=0.2$, $p=2.5$, $\delta=0.01$, and $k=2\pi$. A perturbation of the box size with a small amplitude is introduced in $v_y$ to initiate the instability. The computational domain is covered with either $600\times600$ or $1200\times1200$ grid zones.

The linear growth of KH instability can be analyzed with the linearized version of the RHD equations in (\ref{eq:De}) - (\ref{eq:Ee}). The growth rate was obtained, for instance, in \citet{bodo2004}, for the case of continuous density across the interface, i.e., $\rho_1=\rho_2$. Following the same procedure, it can be derived for the case with a density jump across the interface, such as the one above. Denoting the perturbed quantities as $\delta{\tilde q}\propto\exp[i(kx-\omega t)]$, the dispersion relation is 
\begin{equation}
\begin{aligned}
    \left(\frac{\omega}{k}\right)^6\left(\mathcal{R}_{-}+\frac{U^2}{c^4}\mathcal{C}_-\right)
    +2\left(\frac{\omega}{k}\right)^5U\left[\mathcal{R}_{+}+\left(\frac{1}{c^2}-2\frac{U^2}{c^4}\right)\mathcal{C}_+\right]
    -\left(\frac{\omega}{k}\right)^4U^{2}\left[\mathcal{R}_{-}-\left(\frac{1}{U^2}-\frac{8}{c^2}+6\frac{U^2}{c^4}\right)\mathcal{C}_-\right]~~~\\
    -4\left(\frac{\omega}{k}\right)^3U^3\left[\mathcal{R}_{+}+\left(\frac{1}{U^2}-\frac{3}{c^2}+\frac{U^2}{c^4}\right)\mathcal{C}_+\right]
    -\left(\frac{\omega}{k}\right)^2U^{4}\left[\mathcal{R}_{-}-\left(\frac{6}{U^2}-\frac{8}{c^2}+\frac{U^2}{c^4}\right)\mathcal{C}_-\right]~~~~~~~~~~~~~~~~\\
    +2\left(\frac{\omega}{k}\right)U^{5}\left[\mathcal{R}_{+}-\left(\frac{2}{U^2}-\frac{1}{c^2}\right)\mathcal{C}_+\right]
    +U^{6}\left(\mathcal{R}_{-}+\frac{1}{U^2}\mathcal{C}_-\right)=0,~~~~~~~~~~~~~~~~~~~~~~~~~~~~~~~
\end{aligned}
\label{linearKH}
\end{equation}
where $\mathcal{R} = (\rho_{1}h_{1}c_{s,1})/(\rho_{2}h_{2}c_{s,2})$, $\mathcal{R}_{\pm}=1\pm \mathcal{R}^2$, and $\mathcal{C}_{\pm}=c_{s,2}^{2}\mathcal{R}^2\pm c_{s,1}^{2}$. Here, the subscripts 1 and 2 mark the quantities with the density $\rho_1$ and $\rho_2$, and the pressure is continuous across the interface.

The left panel of Figure \ref{fig:f7} compares the growth of perturbation in the linear phase as measured by the root-mean-square (rms) $y$-velocity, $\left<v_y^2\right>^{1/2}$, with the predicted growth of the $k=2\pi$ mode from Equation (\ref{linearKH}). After the initial adjustment, the simulations show an exponential growth before saturation reaches. Our test demonstrates that with the resolutions employed, the simulations closely reproduce the predicted growth of the linear analysis, regardless of whether the transverse-flux averaging is included or not.

The right panel of Figure \ref{fig:f7} presents the rest-mass density from the simulations without and with the transverse-flux averaging, as well as those using two different grid resolutions; the results are presented at $t=5$ (the end of the linear stage) for the simulations with $600\times600$ grid zones, and  at $t=6.5$ (the nonlinear stage) for the simulations with $1200\times1200$ grid zones, in order to depict the differences caused by the transverse-flux averaging at different evolution phases. Overall, the eddy structures of KH vortices, especially on small scales, better develop by including the transverse-flux averaging. Such structures commonly emerge in astrophysical flows, for instance, in relativistic jets. Hence, we incorporate the transverse-flux averaging as a default component of our code.

\begin{figure*}[t]
\vskip 0.2 cm
\hskip -0.1 cm
\centerline{\includegraphics[width=0.6\linewidth]{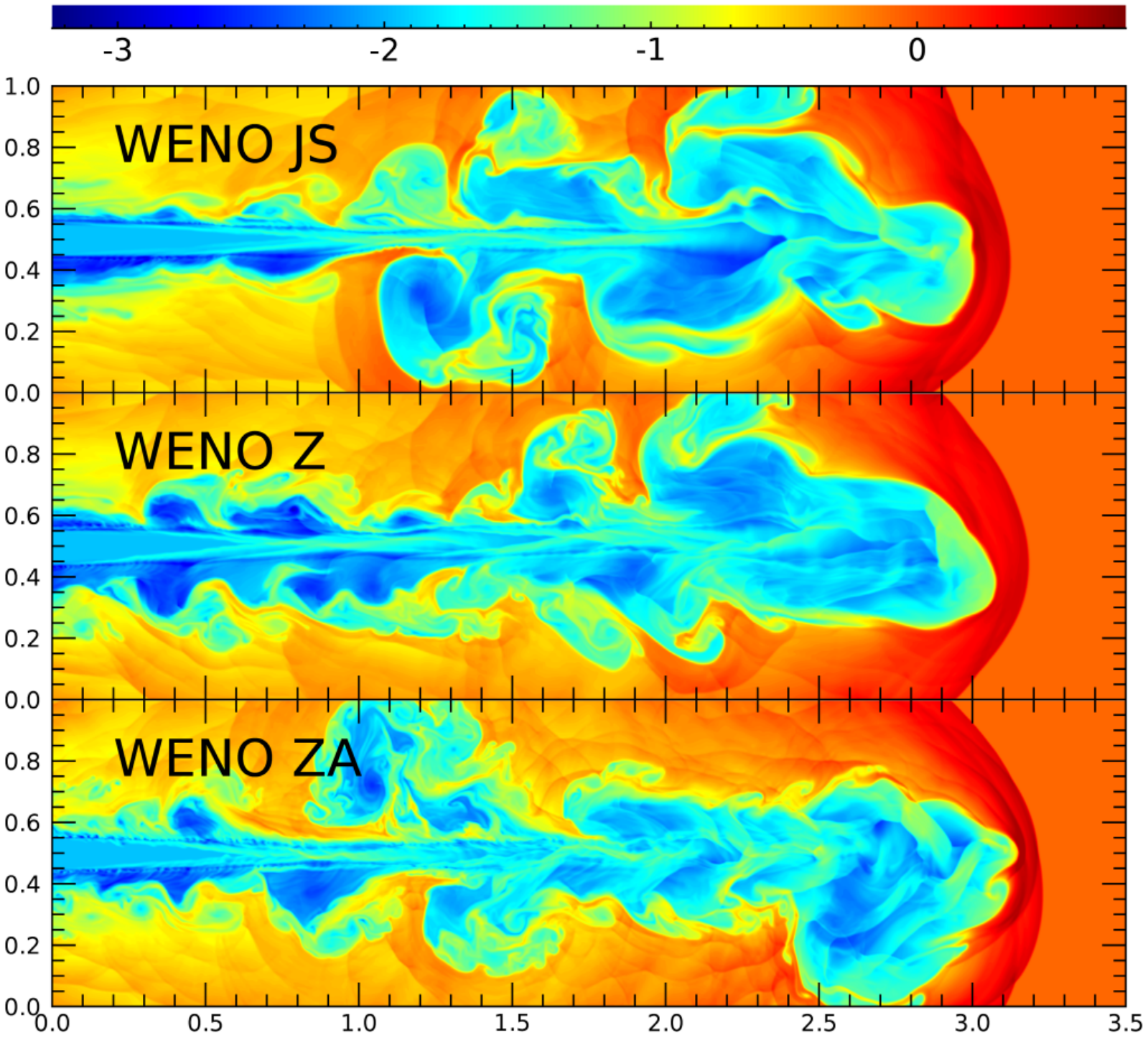}}
\vskip 0.0 cm
\caption{2D plane-parallel (slab) jet with the Lorentz factor $\Gamma_{\rm jet}=7$, injected from the left boundary into a uniform background. The jet and background states are given in Equation (\ref{reljet2d}). The computational domain consists of $[0,3.5]\times [0,1]$ with $1050\times300$ grid zones, and the width of the jet nozzle is 0.1. The rest-mass density, $\log\rho$, from simulations using different WENO versions, WENO-JS (top), WENO-Z (middle), and WENO-ZA (bottom), is shown. Otherwise, the default code setup with SSPRK, RC, CFL = 0.8, and the transverse-flux
averaging is adopted.}
\label{fig:f8}
\end{figure*}

\subsection{Relativistic Jets}
\label{s3.3}

As the main purpose of this paper is to describe a RHD code developed for studies of ultra-relativistic astrophysical jets, we have extensively performed test simulations of jets with different EOSs and different WENO versions as well as different setups for the components of the code. Only some of them are presented here. We point that jet is one of the most popular test problems for RHD codes, and hence many papers for RHD codes presented jet tests (see the introduction for the references).

\subsubsection{2D Plane-Parallel Jet}
\label{s3.3.1}

We first show a test of 2D plan-parallel (slab) jet. The structures of jet and cocoon materials can be visualized more easily with 2D jets than with 3D jets. Simulations have been performed with a light relativistic jet, injected horizontally into the uniform background medium, in a pressure equilibrium:
\begin{equation}
\begin{cases} 
\mbox{\boldmath$u$}_{\rm jet} =(10^{-2},0.99,0,10^{-3}), \\
\mbox{\boldmath$u$}_{\rm bkg} =(1,0,0,10^{-3}), \end{cases}
\label{reljet2d}
\end{equation}
where $\mbox{\boldmath$u$}_{\rm jet}$ and $\mbox{\boldmath$u$}_{\rm bkg}$ denote the jet and background states, respectively. The jet speed of $v_{\rm jet}=0.99$ corresponds to the Lorentz factor of $\Gamma_{\rm jet}=7$. The computational domain covers $[0,3.5]\times [0,1]$ with $1050\times300$ grid zones. A jet nozzle with the width (diameter) of 0.1 is placed at $(x,y)=(0.0,0.5)$. The boundary condition is inflow at the jet nozzle, and outflow elsewhere.

Figure \ref{fig:f8} compares the results of simulations with different WENO versions. In all three cases, the convergence-divergence structure inside the jet spine and the turbulent structure inside the cocoon are well reproduced. With WENO-ZA, clearly more structures develop; however, feather-like patterns appear the interface of the jet-backflow and also in the backflow, and it is not clear whether they are all physical. Between WENO-JS and WENO-Z, the latter produces richer structures. The results of our simulations are favorably compared with those of previous studies, for example, Figure 8 of \citet{nunez2016}, although the jet parameters as well as the EOS in other studies may not be necessarily identical to ours.

\begin{figure*}[t]
\vskip 0.2 cm
\centerline{\includegraphics[width=0.7\linewidth]{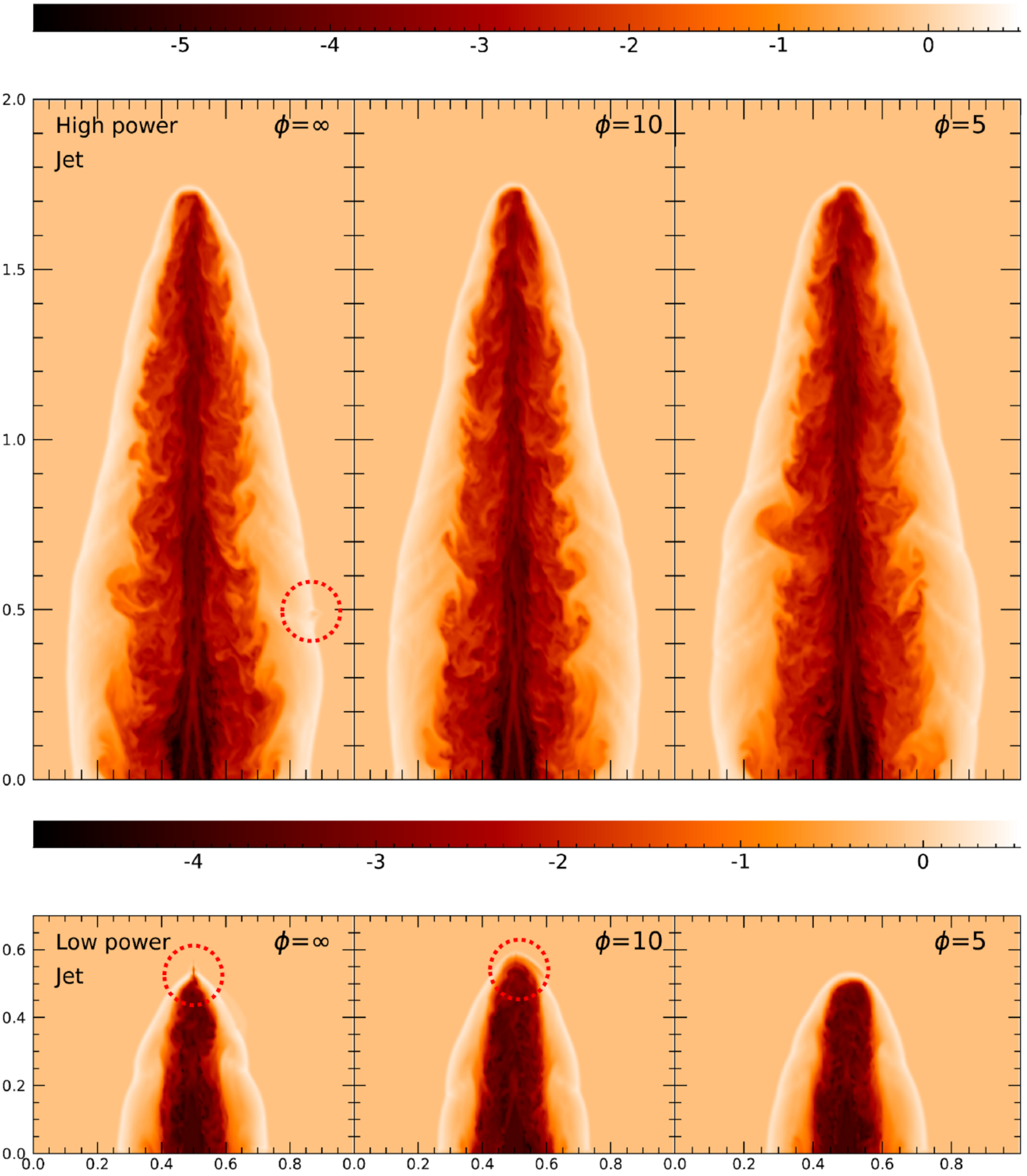}}
\vskip 0.0 cm
\caption{3D jet simulations to demonstrate the suppression of the carbuncle instability with the modification of eigenvalues described in Equation (\ref{carb_cure}). The red circles enclose the regions where the instability starts to appear. The left panels show the simulations {\it without} the modification of eigenvalues, while the center and the right panels are specified {\it with} the modification parameter of $\phi = 10$ and 5, respectively. The top panels display a high-power jet of the Lorentz factor of $\Gamma_{\rm jet}=71$, while the bottom panels show a low-power jet of the Lorentz factor of $\Gamma_{\rm jet}=4.3$. The jet is injected from the center of the bottom $x-y$ plane into a uniform background. The computational domain of $[0,1]\times[0,1]\times[0,2]$ and $[0,1]\times[0,1]\times[0,1]$ is covered with $200\times200\times400$ and $200\times200\times200$ grid zones for the high-power and low-power jets, and the radius of the jet nozzle is 0.03 in both the jets. See the main text for the details of the jet setups. The slice images of $\log_{10}\rho$ in the $x-z$ plane through $y=0.5$ are shown at $t_{\rm end}=50~t_{\rm cross}$ (top) and $13~t_{\rm cross}$ (bottom). The bottom panels include a cropped region of $0\leq z\leq 0.7$. The default code setup with WENO-Z, SSPRK, RC, CFL = 0.8, and the transverse-flux averaging is adopted.}
\label{fig:f9}
\end{figure*}

\begin{figure*}[t]
\vskip 0.2 cm
\centerline{\includegraphics[width=0.9\linewidth]{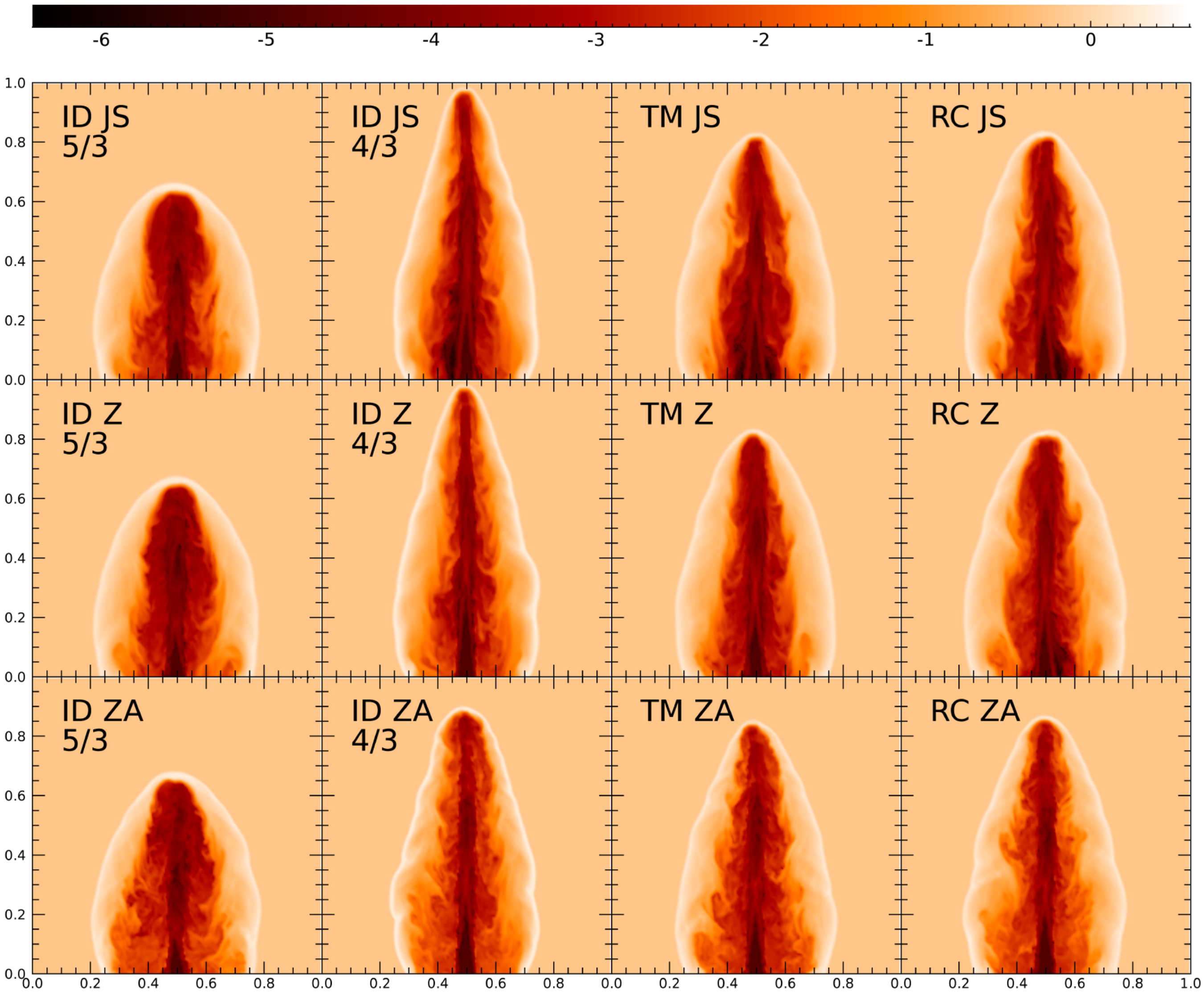}}
\vskip 0.0 cm
\caption{3D jet of the Lorentz factor, $\Gamma_{\rm jet}=71$, injected from the center of the bottom $x-y$ plane into a uniform background. The computational domain of $[0,1]\times[0,1]\times[0,1]$ is covered with $200\times200\times200$ grid zones, and the radius of the jet nozzle is 0.03. See the main text for the details of the jet setup. The slice images of $\log_{10}\rho$ in the $x-z$ plane through $y=0.5$ from simulations with different EOSs, ID with $\gamma=5/3$ (1st column), ID with $\gamma=4/3$ (2nd column), TM (3rd column), and RC (4th column), using different WENO versions, WENO-JS (top), WENO-Z (middle), and WENO-ZA (bottom), are shown at at $t_{\rm end}=25~t_{\rm cross}$. In all the simulations, the transverse-flux averaging and the fix for the carbuncle instability with $\phi = 10$ are implemented, and the SSPRK time integration method with CFL = 0.8 is used.}
\label{fig:f30}
\end{figure*}

\subsubsection{Carbuncle Instability in 3D Jet Simulations}
\label{s3.3.2}

As described in Section \ref{s2.6}, in order to suppress the carbuncle instability at shocks, our code includes the modification of eigenvalues for the acoustic modes, suggested by \citet{fleischmann2020}. We here present 3D jet simulations to demonstrate that the carbuncle instability in the jet is effectively suppressed with this modification.

The morphology of radio galaxy jets, including the FR-I and FR-II dichotomy, is known to depend mainly on the energy injection rate, called the jet power,
\begin{equation}
Q_{\rm jet}=\pi r_{\rm jet}^2v_{\rm jet} \left(\Gamma_{\rm jet}^2 \rho_{\rm jet} h_{\rm jet} - \Gamma_{\rm jet} \rho_{\rm jet} \right),
\end{equation}
defined with the jet quantities including the jet radius, $r_{\rm jet}$ \citep[see, e.g.,][Paper II]{kaiser1997,godfrey2013}. 
We have found that in simulations of high-power jets, the head advances fast and the bow shock surface is relatively stable to the carbuncle instability. By contrast, in low-power jets, the head strolls rather slowly, and the bow shock surface is more prone to the instability.

To illustrate this, we show simulations for two cases representing high-power and low-power jets. 3D jets are set up with the RHD state, $\mbox{\boldmath$u$} = (\rho,v_x,v_y,v_z,p)$, as follows:
\begin{equation}
\begin{cases} 
\mbox{\boldmath$u$}_{\rm jet} =(\eta,v_{\rm jet}\sin(\theta_{\rm pre})\cos(2\pi t/t_{\rm pre}),v_{\rm jet}\sin(\theta_{\rm pre})\sin(2\pi t/t_{\rm pre}), v_{\rm jet}\cos(\theta_{\rm pre}),p), \\
\mbox{\boldmath$u$}_{\rm bkg} =(1,0,0,0,p). \end{cases}
\label{jetcar_1}
\end{equation}
For the high-power jet, the density contract is set to be $\eta=10^{-5}$ and the jet speed is $v_{\rm jet}=0.9999$ (or $\Gamma_{\rm jet}=71$); for the low-power jet, $\eta=10^{-4}$ and $v_{\rm jet}=0.9729$ (or $\Gamma_{\rm jet}=4.3$). The computational domain consists of $[0,1]\times[0,1]\times[0,2]$ for the high-power jet and $[0,1]\times[0,1]\times[0,1]$ for the low-power jet, respectively. The radius of jet nozzle is $r_{\rm jet}=0.03$ for both the cases.

The pressure is set to be $p=7.64\times10^{-6}$ in both the cases, which corresponds to that of the typical ICM with the hydrogen number density, $n_{\rm H,ICM} = 10^{-3}$ cm$^{-3}$, and the temperature, $T_{\rm ICM} = 5\times10^7$ K. Then, assuming that the jet radius is $r_{\rm jet}=1$ kpc, the jet power is $Q_{\rm jet}=3.34\times10^{47}$ ergs s$^{-1}$ and $3.34\times10^{45}$ ergs s$^{-1}$ for the high-power and low-power jets, respectively. Note that $Q_{\rm jet}$ for the high-power jet is relevant for FR-II jets, while $Q_{\rm jet}$ for the low-power jet lies at the border dividing the FR-I and FR-II dichotomy \citep[e.g.,][]{godfrey2013}.

In both the cases of high-power and low-power jets, to break the rotational symmetry, a slow, small-angle precession is added to the jet velocity, $v_{\rm jet}$, with the precession angle and period, $\theta_{\rm pre}=0.5^{\circ}$ and $t_{\rm pre} = 10~t_{\rm cross}$. To ensure a smooth start-up of the jets, preventing possible developments of unphysical structures at the tip of the jets in the early stage, a windowing,
\begin{equation}
v^*_{\rm jet} = v_{\rm jet}\left[1-(r/r_{\rm jet})^n\right]~~{\rm with}~~n=20,
\end{equation}
is applied up to $t=1~t_{\rm cross}$, where $r$ is the radial distance from the jet center. Here, $t_{\rm cross}$ is the jet crossing time defined as $t_{\rm cross} \equiv r_{\rm jet}/v_{\rm head}^*$, where $v_{\rm head}^*$ is the advance speed of the jet head, approximately estimated assuming the balance between the jet ram pressure and the background pressure \citep[][Paper II]{marti1997}. The dynamical evolution of jet-induced flows can be characterized with $t_{\rm cross}$ (see paper II).

Figure \ref{fig:f9} shows the rest-mass density, $\rho$, at $t_{\rm end}=50~t_{\rm cross}$ for the high-power jet and at $t_{\rm end}=13~t_{\rm cross}$ for the low-power jet, from simulations using $200\times200\times400$ and $200\times200\times200$ grid zones, respectively. From the left to right panels, the cases with increasingly greater degrees of the modification of eigenvalues described in Equation (\ref{carb_cure}), $\phi=\infty$, 10, and 5, are shown. Without the modification ($\phi=\infty$), the carbuncle instability appears at the bow shock in both the cases, and will eventually disrupt the jets, if the simulations run longer. The figure illustrates that the instability is suppressed, if the modifications with $\phi=10$ and 5 are applied to the high-power and low-power jets, respectively. The structures in the jet spine, backflow, and shocked ICM, on the other hand, are not much affected by the modification. In general, with our code, the instability is effectively suppressed in simulations with $Q_{\rm jet}\gtrsim10^{46}$ ergs s$^{-1}$ if $\phi\simeq10$ is set, while $\phi\simeq5$ is required in simulations with $Q_{\rm jet}\lesssim10^{46}$ ergs s$^{-1}$.

\subsubsection{3D Relativistic Jet}
\label{s3.3.3}

As the final test, we present 3D simulations of a high-power FR-II jet with different EOSs and different WENO versions. The jet parameters are the same as those of the high-power jet in the upper panels of Figure \ref{fig:f9}. The computational domain is $[0,1]\times[0,1]\times[0,1]$ with $200\times200\times200$ grid zones, smaller than that of Figure \ref{fig:f9}. Figure \ref{fig:f30} shows the rest-mass density, $\rho$, at $t_{\rm end}=25~t_{\rm cross}$ from simulations with 12 combinations of different EOSs and different WENO versions. The modification of eigenvalues with $\phi=10$ is included. When the upward-moving jet penetrates into the ICM, the jet flow is halted at the jet head, and reflected backward, producing a cocoon of downward-moving backflow with the low-density jet material. In addition, a bow shock forms in the ICM, encompassing the shocked ICM and the jet cocoon. The KH instability commonly develops at the interface between the jet spine and backflow and also at the interface between the backflow and shocked ICM.

Figure \ref{fig:f30} demonstrates that the advance of the jet and also the overall morphology depend on the EOS. The structures developed in the simulations with RC and TM are comparable. However, the results using ID, with both $\gamma=5/3$ and 4/3, look sufficiently different. In fact, in the simulations with RC and TM, while the injected jet material has the adiabatic index $\gamma\simeq1.45$, it varies from $\gamma\simeq4/3$ at the shocked jet flow to $\gamma\simeq5/3$ at the shocked ICM. Inspecting closely the figure reveals that WENO-JS is the most diffusive among the three WENO-versions, while WENO-ZA is the least diffusive and produces the most complex structures. However, as noted above (e.g. Figures \ref{fig:f5} - \ref{fig:f6}), WENO-ZA sometimes produces suspicious features. Considering both the accuracy and robustness, we hence employ WENO-Z as the default WENO reconstruction scheme for our RHD code.

\section{Summary and Discussion}
\label{s4}

To study the properties of ultra-relativistic astrophysical jets with numerical simulations, we have developed a RHD code based on the 5th-order accurate FD WENO scheme \citep{jiang1996}. As an effort to find an optimal code to emulate accurately and robustly  the nonlinear structures and flow dynamics as well as the thermodynamics in the jets, we have incorporated several recent developments available in the literature: (1) ID, TM, and RC EOSs, (2) WENO-JS, WENO-Z, and WENO-ZA versions, (3) RK4 and SSPRK time integration methods, (4) a high-order accurate averaging of fluxes along the transverse directions, and (5) a modification of eigenvalues for the acoustic modes to suppress the carbuncle instability.

We then have performed an extensive set of test simulations for 1D, 2D, and 3D problems, to evaluate the effectiveness of these implementations and hence to identify the best combination of different options for the studies of relativistic jets. 
In conclusion, we have chosen the RC EOS, WENO-Z scheme, and SSPRK integration method as the fiducial setup of our code. In addition, we include the 4th-order accurate averaging of fluxes along the transverse directions and the modification of eigenvalues for the acoustic modes suggested by \citet{fleischmann2020} in the code.

The main points of this paper can be summarized as follows:

1. In simulations of ultra-relativistic jets, in which the jet of $\gamma\simeq4/3$ with relativistic temperature penetrates into the ICM of $\gamma\simeq5/3$ with subrelativistic temperature, the two fluids are mixed by turbulence in the cocoon, producing the transition from $\gamma\simeq4/3$ to $\gamma\simeq5/3$. In order to follow such process, either the RC or TM EOS should be used, instead of the ID EOS with a fixed $\gamma$. We find RC and TM produce comparable results in various tests presented here. Yet, considering that RC approximates RP more closely than TM \citep{ryu2006}, we adopt RC as the default EOS.

2. Among the three versions of WENO schemes considered here, WENO-JS is the most diffusive, while WENO-ZA is the least diffusive, as expected. WENO-ZA tends to produce richer structures, but at the same time seems to be less robust under harsh conditions such as those of relativistic jets. Considering that WENO-Z is sufficiently accurate and also robust, it is selected as the default option to be used for our simulations of relativistic jets.

3. Between the two 4th-order accurate time integration methods, SSPRK yields better results than RK4 for simulations of shocks with large transverse velocities. We hence adopt SSPRK as the default time integration method.

4. Borrowing the idea of the modified dimension-by-dimension method suggested for FV WENO schemes \citep{buchmuller2014,buchmuller2016}, we implement a 4th-order accurate averaging of fluxes along the transverse directions in smooth flow regions. It improves the performance in some multi-dimensional problems, especially in those involving complex flows.

5. In our simulations of relativistic jets, the bow shock surface is frequently subject to the carbuncle instability. We find that the modification of eigenvalues for the acoustic modes described in Equation (\ref{carb_cure}) effectively suppresses the instability with the parameter, $\phi\sim 5-10$.

In an accompanying paper (Paper II), using the RHD code presented here, we report a simulation study of ultra-relativistic jets ejected from AGNs into the uniform ICM. In particular, we describe in detail the flow dynamics, such as shocks, velocity shear, turbulence, induced by the jets.

\begin{acknowledgments}
We thank the anonymous referee for constructive comments on the manuscript. This work was supported by the National Research Foundation (NRF) of Korea through grants 2016R1A5A1013277, 2020R1A2C2102800, and 2020R1F1A1048189. The work of J.S. was also supported by the NRF through grant 2020R1A6A3A13071702. Some of simulations were performed using the high performance computing resources of the UNIST Supercomputing Center.
\end{acknowledgments}

\bibliography{RHDcode}{}

\begin{thebibliography}{}
\expandafter\ifx\csname natexlab\endcsname\relax\def\natexlab#1{#1}\fi
\providecommand{\url}[1]{\href{#1}{#1}}
\providecommand{\dodoi}[1]{doi:~\href{http://doi.org/#1}{\nolinkurl{#1}}}
\providecommand{\doeprint}[1]{\href{http://ascl.net/#1}{\nolinkurl{http://ascl.net/#1}}}
\providecommand{\doarXiv}[1]{\href{https://arxiv.org/abs/#1}{\nolinkurl{https://arxiv.org/abs/#1}}}

\bibitem[{{Aloy} {et~al.}(1999){Aloy}, {Ib{\'a}{\~n}ez}, {Mart{\'\i}}, \&
  {M{\"u}ller}}]{aloy1999}
{Aloy}, M.~A., {Ib{\'a}{\~n}ez}, J.~M., {Mart{\'\i}}, J.~M., \& {M{\"u}ller},
  E. 1999, \apjs, 122, 151, \dodoi{10.1086/313214}

\bibitem[{{Bodo} {et~al.}(2004){Bodo}, {Mignone}, \& {Rosner}}]{bodo2004}
{Bodo}, G., {Mignone}, A., \& {Rosner}, R. 2004, \pre, 70, 036304,
  \dodoi{10.1103/PhysRevE.70.036304}

\bibitem[{Borges {et~al.}(2008)Borges, Carmona, Costa, \& Don}]{borges2008}
Borges, R., Carmona, M., Costa, B., \& Don, W.~S. 2008, J. Comput. Phys., 227,
  3191, \dodoi{10.1016/j.jcp.2007.11.038}

\bibitem[{Buchmüller {et~al.}(2016)Buchmüller, Dreher, \&
  Helzel}]{buchmuller2016}
Buchmüller, P., Dreher, J., \& Helzel, C. 2016, Applied Mathematics and
  Computation, 272, 460, \dodoi{10.1016/j.amc.2015.03.078}

\bibitem[{{Buchm{\"u}ller} \& {Helzel}(2014)}]{buchmuller2014}
{Buchm{\"u}ller}, P., \& {Helzel}, C. 2014, Journal of Scientific Computing,
  61, \dodoi{10.1007/s10915-014-9825-1}

\bibitem[{{Chattopadhyay} \& {Ryu}(2009)}]{chattopadhyay2009}
{Chattopadhyay}, I., \& {Ryu}, D. 2009, \apj, 694, 492,
  \dodoi{10.1088/0004-637X/694/1/492}

\bibitem[{{Del Zanna} \& {Bucciantini}(2002)}]{del2002}
{Del Zanna}, L., \& {Bucciantini}, N. 2002, \aap, 390, 1177,
  \dodoi{10.1051/0004-6361:20020776}

\bibitem[{Dolezal \& Wong(1995)}]{dolezal1995}
Dolezal, A., \& Wong, S. 1995, J. Comput. Phys., 120, 266

\bibitem[{{Duffell} \& {MacFadyen}(2011)}]{duffell2011}
{Duffell}, P.~C., \& {MacFadyen}, A.~I. 2011, \apjs, 197, 15,
  \dodoi{10.1088/0067-0049/197/2/15}

\bibitem[{Dumbser {et~al.}(2004)Dumbser, Moschetta, \& Gressier}]{dumbser2004}
Dumbser, M., Moschetta, J.-M., \& Gressier, J. 2004, J. Comput. Phys., 197,
  647, \dodoi{10.1016/j.jcp.2003.12.013}

\bibitem[{{Falle} \& {Komissarov}(1996)}]{falle1996}
{Falle}, S.~A.~E.~G., \& {Komissarov}, S.~S. 1996, \mnras, 278, 586,
  \dodoi{10.1093/mnras/278.2.586}

\bibitem[{Fleischmann {et~al.}(2020)Fleischmann, Adami, Hu, \&
  Adams}]{fleischmann2020}
Fleischmann, N., Adami, S., Hu, X.~Y., \& Adams, N.~A. 2020, J. Comput. Phys.,
  401, 109004, \dodoi{10.1016/j.jcp.2019.109004}

\bibitem[{{Gaensler} \& {Slane}(2006)}]{gaensler2006}
{Gaensler}, B.~M., \& {Slane}, P.~O. 2006, \araa, 44, 17,
  \dodoi{10.1146/annurev.astro.44.051905.092528}

\bibitem[{{Godfrey} \& {Shabala}(2013)}]{godfrey2013}
{Godfrey}, L.~E.~H., \& {Shabala}, S.~S. 2013, \apj, 767, 12,
  \dodoi{10.1088/0004-637X/767/1/12}

\bibitem[{Gottlieb(2005)}]{gottlieb2005}
Gottlieb, S. 2005, Journal of Scientific Computing, 25, 105,
  \dodoi{10.1007/BF02728985}

\bibitem[{Ha {et~al.}(2013)Ha, Kim, Lee, \& Yoon}]{ha2013}
Ha, Y., Kim, C.~H., Lee, Y.~J., \& Yoon, J. 2013, J. Comput. Phys., 232, 68,
  \dodoi{10.1016/j.jcp.2012.06.016}

\bibitem[{{Hanawa} {et~al.}(2008){Hanawa}, {Mikami}, \&
  {Matsumoto}}]{hanawa2008}
{Hanawa}, T., {Mikami}, H., \& {Matsumoto}, T. 2008, in Astronomical Society of
  the Pacific Conference Series, Vol. 385, Numerical Modeling of Space Plasma
  Flows, ed. N.~V. {Pogorelov}, E.~{Audit}, \& G.~P. {Zank}, 259

\bibitem[{{Hardcastle} \& {Croston}(2020)}]{hardcastle2020}
{Hardcastle}, M.~J., \& {Croston}, J.~H. 2020, arXiv e-prints,
  arXiv:2003.06137.
\newblock \doarXiv{2003.06137}

\bibitem[{He \& Tang(2012)}]{he2012}
He, P., \& Tang, H. 2012, Communications in Computational Physics, 11,
  114–146, \dodoi{10.4208/cicp.291010.180311a}

\bibitem[{Henrick {et~al.}(2005)Henrick, Aslam, \& Powers}]{henrick2005}
Henrick, A.~K., Aslam, T.~D., \& Powers, J.~M. 2005, J. Comput. Phys., 207,
  542, \dodoi{10.1016/j.jcp.2005.01.023}

\bibitem[{Hu {et~al.}(2010)Hu, Wang, \& Adams}]{hu2010}
Hu, X., Wang, Q., \& Adams, N.~A. 2010, J. Comput. Phys., 229, 8952,
  \dodoi{10.1016/j.jcp.2010.08.019}

\bibitem[{Jiang \& Shu(1996)}]{jiang1996}
Jiang, G.-S., \& Shu, C.-W. 1996, J. Comput. Phys., 126, 202,
  \dodoi{10.1006/jcph.1996.0130}

\bibitem[{Jiang \& Wu(1999)}]{jiang1999}
Jiang, G.-S., \& Wu, C.-C. 1999, J. Comput. Phys., 150, 561,
  \dodoi{10.1006/jcph.1999.6207}

\bibitem[{{Joshi} {et~al.}(2021){Joshi}, {Chattopadhyay}, {Ryu}, \&
  {Yadav}}]{josh2021}
{Joshi}, R.~K., {Chattopadhyay}, I., {Ryu}, D., \& {Yadav}, L. 2021, \mnras,
  502, 5227, \dodoi{10.1093/mnras/stab364}

\bibitem[{{Kaiser} \& {Alexander}(1997)}]{kaiser1997}
{Kaiser}, C.~R., \& {Alexander}, P. 1997, \mnras, 286, 215,
  \dodoi{10.1093/mnras/286.1.215}

\bibitem[{{Landau} \& {Lifshitz}(1959)}]{landau1959}
{Landau}, L.~D., \& {Lifshitz}, E.~M. 1959, {Fluid mechanics}

\bibitem[{{Lister} {et~al.}(2013){Lister}, {Aller}, {Aller}, {Homan},
  {Kellermann}, {Kovalev}, {Pushkarev}, {Richards}, {Ros}, \&
  {Savolainen}}]{lister2013}
{Lister}, M.~L., {Aller}, M.~F., {Aller}, H.~D., {et~al.} 2013, \aj, 146, 120,
  \dodoi{10.1088/0004-6256/146/5/120}

\bibitem[{Liu {et~al.}(2018)Liu, Shen, Zeng, \& Yu}]{liu2018}
Liu, S., Shen, Y., Zeng, F., \& Yu, M. 2018, International Journal for
  Numerical Methods in Fluids, 87, 271, \dodoi{10.1002/fld.4490}

\bibitem[{Liu {et~al.}(1994)Liu, Osher, \& Chan}]{liu1994}
Liu, X.-D., Osher, S., \& Chan, T. 1994, J. Comput. Phys., 115, 200,
  \dodoi{10.1006/jcph.1994.1187}

\bibitem[{Mart{\i} \& M{\"u}ller(1996)}]{marti1996}
Mart{\i}, J.~M., \& M{\"u}ller, E. 1996, J. Comput. Phys., 123, 1

\bibitem[{{Mart{\'\i}} \& {M{\"u}ller}(2003)}]{marti2003}
{Mart{\'\i}}, J.~M., \& {M{\"u}ller}, E. 2003, Living Reviews in Relativity, 6,
  7, \dodoi{10.12942/lrr-2003-7}

\bibitem[{{Mart{\'\i}} \& {M{\"u}ller}(2015)}]{marti2015}
---. 2015, Living Reviews in Computational Astrophysics, 1, 3,
  \dodoi{10.1007/lrca-2015-3}

\bibitem[{Mart{\'\i} {et~al.}(1997)Mart{\'\i}, M{\"u}ller, Font,
  Ib{\'a}{\~n}ez, \& Marquina}]{marti1997}
Mart{\'\i}, J.-M., M{\"u}ller, E., Font, J.~A., Ib{\'a}{\~n}ez, J.-M., \&
  Marquina, A. 1997, \apj, 479, 151, \dodoi{10.1086/303842}

\bibitem[{{Mathews}(1971)}]{mathews1971}
{Mathews}, W.~G. 1971, \apj, 165, 147, \dodoi{10.1086/150883}

\bibitem[{{Mignone} \& {Bodo}(2005)}]{mignone2005a}
{Mignone}, A., \& {Bodo}, G. 2005, \mnras, 364, 126,
  \dodoi{10.1111/j.1365-2966.2005.09546.x}

\bibitem[{{Mignone} {et~al.}(2005){Mignone}, {Plewa}, \& {Bodo}}]{mignone2005b}
{Mignone}, A., {Plewa}, T., \& {Bodo}, G. 2005, \apjs, 160, 199,
  \dodoi{10.1086/430905}

\bibitem[{{Nava} {et~al.}(2017){Nava}, {Desiante}, {Longo}, {Celotti},
  {Omodei}, {Vianello}, {Bissaldi}, \& {Piran}}]{nava2017}
{Nava}, L., {Desiante}, R., {Longo}, F., {et~al.} 2017, \mnras, 465, 811,
  \dodoi{10.1093/mnras/stw2771}

\bibitem[{{N{\'u}{\~n}ez-de la Rosa} \& {Munz}(2016)}]{nunez2016}
{N{\'u}{\~n}ez-de la Rosa}, J., \& {Munz}, C.-D. 2016, \mnras, 460, 535,
  \dodoi{10.1093/mnras/stw999}

\bibitem[{Pandolfi \& D'Ambrosio(2001)}]{pandolfi2001}
Pandolfi, M., \& D'Ambrosio, D. 2001, J. Comput. Phys., 166, 271,
  \dodoi{10.1006/jcph.2000.6652}

\bibitem[{Peery \& Imlay(1988)}]{peery1988}
Peery, K., \& Imlay, S. 1988, in 24th Joint Propulsion Conference, 2904,
  \dodoi{10.2514/6.1988-2904}

\bibitem[{Piran(2005)}]{piran2005}
Piran, T. 2005, Rev. Mod. Phys., 76, 1143, \dodoi{10.1103/RevModPhys.76.1143}

\bibitem[{{Pons} {et~al.}(2000){Pons}, {Ma Mart{\'\i}}, \&
  {M{\"u}ller}}]{pons2000}
{Pons}, J.~A., {Ma Mart{\'\i}}, J., \& {M{\"u}ller}, E. 2000, Journal of Fluid
  Mechanics, 422, 125, \dodoi{10.1017/S0022112000001439}

\bibitem[{Qin {et~al.}(2016)Qin, Shu, \& Yang}]{qin2016}
Qin, T., Shu, C.-W., \& Yang, Y. 2016, J. Comput. Phys., 315, 323

\bibitem[{{Radice} \& {Rezzolla}(2012)}]{radice2012}
{Radice}, D., \& {Rezzolla}, L. 2012, \aap, 547, A26,
  \dodoi{10.1051/0004-6361/201219735}

\bibitem[{Reyes {et~al.}(2019)Reyes, Lee, Graziani, \& Tzeferacos}]{reyes2019}
Reyes, A., Lee, D., Graziani, C., \& Tzeferacos, P. 2019, J. Comput. Phys.,
  381, 189

\bibitem[{{Ryu} {et~al.}(2006){Ryu}, {Chattopadhyay}, \& {Choi}}]{ryu2006}
{Ryu}, D., {Chattopadhyay}, I., \& {Choi}, E. 2006, \apjs, 166, 410,
  \dodoi{10.1086/505937}

\bibitem[{{Savolainen} {et~al.}(2010){Savolainen}, {Homan}, {Hovatta},
  {Kadler}, {Kovalev}, {Lister}, {Ros}, \& {Zensus}}]{Savolainen2010}
{Savolainen}, T., {Homan}, D.~C., {Hovatta}, T., {et~al.} 2010, \aap, 512, A24,
  \dodoi{10.1051/0004-6361/200913740}

\bibitem[{Schneider {et~al.}(1993)Schneider, Katscher, Rischke, Waldhauser,
  Maruhn, \& Munz}]{schneider1993}
Schneider, V., Katscher, U., Rischke, D., {et~al.} 1993, J. Comput. Phys., 105,
  92

\bibitem[{{Seo} {et~al.}(2021){Seo}, {Kang}, \& {Ryu}}]{seo2021b}
{Seo}, J., {Kang}, H., \& {Ryu}, D. 2021, \apj, submitted (Paper II,
  arXiv:2106.04100)

\bibitem[{Shu(2009)}]{shu2009}
Shu, C.-W. 2009, SIAM Review, 51, 82, \dodoi{10.1137/070679065}

\bibitem[{Shu \& Osher(1988)}]{shu1988}
Shu, C.-W., \& Osher, S. 1988, J. Comput. Phys., 77, 439,
  \dodoi{10.1016/0021-9991(88)90177-5}

\bibitem[{Shu \& Osher(1989)}]{shu1989}
---. 1989, J. Comput. Phys., 83, 32, \dodoi{10.1016/0021-9991(89)90222-2}

\bibitem[{Spiteri \& Ruuth(2002)}]{spiteri2002}
Spiteri, R.~J., \& Ruuth, S.~J. 2002, SIAM Journal on Numerical Analysis, 40,
  469, \dodoi{10.1137/S0036142901389025}

\bibitem[{Spiteri \& Ruuth(2003)}]{spiteri2003}
---. 2003, Mathematics and Computers in Simulation, 62, 125,
  \dodoi{10.1016/S0378-4754(02)00179-9}

\bibitem[{Synge(1957)}]{synge1957}
Synge, J.~L. 1957, The Relativistic Gas, Series in physics (North-Holland
  Publishing Company)

\bibitem[{Taub(1948)}]{taub1948}
Taub, A.~H. 1948, Phys. Rev., 74, 328, \dodoi{10.1103/PhysRev.74.328}

\bibitem[{Woodward \& Colella(1984)}]{wood1984}
Woodward, P., \& Colella, P. 1984, J. Comput. Phys., 54, 115,
  \dodoi{10.1016/0021-9991(84)90142-6}

\bibitem[{Zhang {et~al.}(2011)Zhang, Zhang, \& Shu}]{zhang2011}
Zhang, R., Zhang, M., \& Shu, C.-W. 2011, Communications in Computational
  Physics, 9, 807–827, \dodoi{10.4208/cicp.291109.080410s}

\bibitem[{{Zhang} \& {MacFadyen}(2006)}]{zhang2006}
{Zhang}, W., \& {MacFadyen}, A.~I. 2006, \apjs, 164, 255,
  \dodoi{10.1086/500792}

\end{thebibliography}
\bibliographystyle{aasjournal}

\end{document}